\documentclass[12pt]{article}
\usepackage{amsmath} 
\usepackage{multirow} 
\usepackage{amsfonts}
\usepackage{amssymb}
\def\hybrid{\topmargin -20pt    \oddsidemargin 0pt
        \headheight 0pt \headsep 0pt
        \textwidth 6.25in       
        \textheight 9.5in       
        \marginparwidth .875in
        \parskip 5pt plus 1pt   \jot = 1.5ex}
\hybrid
\numberwithin{equation}{section}
\numberwithin{table}{section}
\newcommand{\beq}{\begin{equation}}
\newcommand{\eeq}{\end{equation}}
\newcommand{\bi}{\begin{itemize}}
\newcommand{\ei}{\end{itemize}}
\newcommand{\bea}{\begin{eqnarray}}
\newcommand{\eea}{\end{eqnarray}}
\newcommand{\ba}{\begin{array}}
\newcommand{\ea}{\end{array}}
\newcommand{\bt}{\begin{tabular}}
\newcommand{\et}{\end{tabular}}
\newcommand{\bc}{\begin{center}}
\newcommand{\ec}{\end{center}}

\newcommand{\Ox}{\Omega}

\newcommand{\Oxb}{\bar{\Omega}}
\newcommand{\cD}{\mathcal{D}}
\newcommand{\cL}{\mathcal{L}}
\newcommand{\cK}{\mathcal{K}}
\newcommand{\cN}{\mathcal{N}}
\newcommand{\cG}{\mathcal{G}}

\newcommand{\cF}{\mathcal{F}}

\newcommand{\cM}{\mathcal M}

\newcommand{\Vw}{{\mathcal K}_w\vphantom{{\mathcal V}_w}}
\newcommand{\Gw}{{\mathcal G}_w\vphantom{{\mathcal G}_w}}
\newcommand{\ib}{{\bar\imath}}
\newcommand{\jb}{{\bar\jmath}}
\newcommand{\bj}{\bar{\jmath}}
\newcommand{\bk}{\bar{k}}
\newcommand{\bl}{\bar{l}}

\newcommand{\CY}{Calabi--Yau}
\newcommand{\nn}{\nonumber}

\newcommand{\M}{M}
\newcommand{\Weff}{W^{\rm (eff)}}
\newcommand{\Weffb}{\bar W^{\rm (eff)}}
\newcommand{\Y}{Y}
\newcommand{\G}{\mathcal{I}}
\newcommand{\CHI}{\mathcal{I}}

\newcommand{\hW}{\hat{W}}
\newcommand{\hK}{\hat{K}}

\newcommand{\hphi}{{\phi}}
\newcommand{\Gt}{G^{(3)}\vphantom{G}} 
\newcommand{\Gtb}{\bar{G}^{(3)}\vphantom{\bar G}} 
\newcommand{\ha}{\hat{a}}
\newcommand{\hb}{\hat{b}}
\newcommand{\hab}{\hat{ a}}
\newcommand{\hbb}{\hat{ b}}
\newcommand{\cha}{\chi_{\hat{a}}}
\newcommand{\chab}{\bar{\chi}_{\hat{a}}}

\newcommand{\Dth}{\rm (D3)}
\newcommand{\eff}{\rm (eff)}
\newcommand{\tr}{\mathrm{Tr}\:}
\newcommand{\id}{\mathbf{1}}
\newcommand{\com}[2]{\big[ {#1},{#2} \big]}
\newcommand{\lie}[2]{\left[ {#1},{#2} \right]}
\newcommand{\ins}[1]{\mathrm{i}_{#1}}
\newcommand{\D}{\mathrm{D}}        
\newcommand{\Kw}{\mathcal{K}_w}    
\newcommand{\Em}{\varphi}          
\newcommand{\WV}{\mathcal{W}}      
\newcommand{\FD}{F}
\newcommand{\FA}{F_\mathrm{A}}
\newcommand{\norm}[1]{\lVert #1\rVert}
\newcommand{\Riem}[4]{R_{#1\hphantom{#2}#3#4}^{\hphantom{#1}#2}}

\begin{document}

\begin{titlepage}
\begin{center}

\hfill hep-th/0312234\\
\vskip 0.3cm
\rightline{\small CPHT-RR 116.1203}

\vskip 1.5cm

{\large \bf  Soft Supersymmetry Breaking in

 Calabi-Yau Orientifolds
with D-branes and Fluxes}\footnote{%
Work supported by: DFG -- The German Science Foundation,
European RTN Program HPRN-CT-2000-00148 and the
DAAD -- the German Academic Exchange Service.}\\

\vskip 0.8cm

{\bf Mariana Gra{\~n}a}\\

\vskip 0.3cm
{\em Centre de Physique Th{\'e}orique\\
Ecole Polytechnique \\
91128 Palaiseau Cedex, France}\\
{\tt Mariana.Grana@cpht.polytechnique.fr}
\vskip 0.8cm

{\bf Thomas W.\ Grimm, Hans Jockers and Jan Louis}  \\
\vskip 0.3cm

{\em II. Institut f{\"u}r Theoretische Physik\\
Universit{\"a}t Hamburg\\
Luruper Chaussee 149\\
 D-22761 Hamburg, Germany}\\
\vskip 10pt

 {\tt  thomas.grimm, hans.jockers, jan.louis@desy.de} \\

\end{center}

\vskip 1cm

\begin{center} {\bf ABSTRACT } \end{center}

\noindent
In this paper we compute the $\cN=1 $ effective low energy action for a 
stack of $N$ space-time filling D3-branes in generic type IIB
Calabi-Yau orientifolds  with non-trivial  background fluxes
by reducing the Dirac-Born-Infeld and Chern-Simons actions.
Specifically, we determine the K\"ahler potential for
the excitations of the D-brane including their couplings
to all bulk moduli fields.
In the effective theory, $\cN=1 $ supergravity is spontaneously
broken by the presence of fluxes and we compute the
induced soft supersymmetry breaking terms.
We find an interesting structure
in the resulting soft terms with generically universal soft
scalar masses.

\vfill

\noindent December 2003

\end{titlepage}

\section{Introduction}
\setcounter{equation}{0}

After the discovery of D-branes as non-perturbative 
BPS objects in string theory \cite{JP},
it was realized that they also can serve as a
new ingredient in string model building.
Recently phenomenologically viable models
with chiral fermions in 
representations of gauge groups similar to the $SU(3)\times SU(2)\times U(1)$
of the Standard Model (SM) were constructed from 
stacks of space-time filling D-branes \cite{review}.
Non-supersymmetric models generically suffer from 
instabilities and a hierarchy problem while  in 
supersymmetric models 
a viable mechanism for hierarchical supersymmetry
breaking has to be employed.
Apart from standard non-perturbative mechanisms for
supersymmetry breaking such as gaugino condensation,
it has been suggested to use background fluxes
in order to stabilize the moduli and break supersymmetry
spontaneously
\cite{Bachas}--\cite{DWG}.

{}From a phenomenological point of view spontaneously broken $\cN=1$
theories are of particular interest.
Starting from type II string theories in ten space-time dimensions
($D=10$), one can compactify  on Calabi-Yau
threefolds to obtain  $\cN=2$ theories in $D=4$.
This $\cN=2$ is further broken to $\cN=1$ if in addition
background D-branes and/or orientifold planes are present.
Turning on background fluxes results in a 
spontaneously broken  $\cN=1$
theory which can be best
discussed in terms of a low energy effective action.
For Calabi-Yau compactifications of type IIB theories
this effective theory was determined in refs.\
\cite{Michelson,BGHL,DallAgata,LM}  while for 
Calabi-Yau orientifolds it can be found in refs.\ \cite{GKP,BBHL,DWG,GL}.

In brane-world models the SM, or rather its supersymmetric extension,
arises from the dynamical excitations of the D-branes.
Thus it is necessary to determine the low energy effective 
action of these excitations including their couplings to
the gauge neutral bulk moduli fields. 
This can be done in different ways.
In refs.\ \cite{BHK} T-duality is used to determine the effective 
action for a stack of D3-branes starting from the type I action and in refs.\ \cite{Ferrara}
supergravity consistency considerations are employed to 
determine the effective action.\footnote{This works particularly well
for backgrounds with a high degree of supersymmetry, but appears to be less efficient for $\cN=1$ theories.} 
The most direct way is to use the Dirac-Born-Infeld (DBI) 
and Chern-Simons (CS) action 
(and their supersymmetric extension)
in an appropriate bulk background
and perform a Kaluza-Klein reduction \cite{FP,MG,D3}.
It is the latter approach we will follow in this paper.

The presence of background fluxes breaks supersymmetry,
which is communicated to the observable sector, i.e.\
the charged matter fields on the D-branes, by bulk moduli fields.
This results in a set of soft supersymmetry breaking terms
which can be computed from the effective low energy action.
For a specific class of models this has been carried out
in refs.\ \cite{MG,KN}.

The purpose of this paper is twofold.
First (section \ref{sec:actions}), we determine the low energy effective action
of type IIB string theory compactified 
on Calabi-Yau orientifolds with a stack of space-time filling
D-branes and background fluxes. 
For the bulk action (section \ref{subsec:bulk}), 
we truncate the spectrum as dictated by the orientifold
projection, and then reduce the ten-dimensional 
type IIB supergravity action. 
For the D-branes (section \ref{subsec:D3action}), 
we use the bosonic non-Abelian Dirac-Born-Infeld 
and Chern-Simons action as proposed in \cite{Myers}
and their supersymmetric completion as obtained in \cite{D3} 
using the $\kappa$-symmetric action of \cite{superaction}.\footnote{%
For simplicity we confine our analysis to
D3-branes leaving the  higher-dimensional D-branes to a 
separate investigation. }
We determine the K\"ahler potential of the charged matter excitations
of the D-brane coupled to all Calabi-Yau orientifold bulk moduli
including the complex structure deformations. 
We find that generically 
it is not of the `sequestered form' \cite{RS} in agreement with the
discussion of  ref.\ \cite{ADGT}.
In the limit of just one K\"ahler modulus (parameterizing the overall volume)
and frozen
complex structure moduli we confirm the 
K\"ahler potential suggested in refs.\ \cite{DWG, KKLMMT}.

The second purpose of this paper (section \ref{sec:ssb})
is to compute the soft supersymmetry breaking terms arising from 
turning on three-form flux.
After briefly reviewing 
the generic supergravity analysis of refs.\ \cite{KL,BIM}
(section \ref{subsec:sugra}),
we determine the resulting supersymmetric (section \ref{subsec:softD3})
and soft breaking terms (section \ref{subsec:SSB}) in the D-brane action.
We find that the fermionic masses are generated by a Giudice-Masiero
mechanism \cite{GM} induced via fluxes into specific couplings in the 
D-brane K\"ahler potential.
In sections \ref{subsec:ISD} and \ref{subsec:IASD}
we briefly discuss some phenomenological properties  
of the resulting soft supersymmetry breaking terms.
We find that for (0,3) fluxes a `strict no-scale' breaking \cite{NS}
occurs in that all soft terms vanish. For (3,0) and (1,2) fluxes 
on the other hand, we find $A$-terms which
are proportional to the Yukawa couplings, and universal
scalar masses.

We also display 
the consistency of the computed soft terms with the generic
supergravity formulas of refs.\ \cite{KL,BIM}.
This is a highly non-trivial check on the
computation performed in section \ref{sec:actions}.
It is important to stress that we do not choose a particular
model but  our analysis is valid for any Calabi-Yau orientifold
compactifications with D3-branes and background fluxes.

When this manuscript was being prepared 
the paper \cite{CIM} appeared, which has substantial overlap
with our analysis.


\section{Effective Actions}\label{sec:actions}


\subsection{The Bulk: Calabi-Yau orientifolds with three-form flux}\label{subsec:bulk}
In order to set the stage for this paper let us first
briefly summarize the results of ref.\ \cite{GL}
where the low energy effective action of 
Calabi-Yau orientifold compactifications of type IIB string theory
is derived. The class of orientifolds studied are obtained by 
modding out  type IIB string theory by the world-sheet
parity combined with a discrete isometry of the Calabi-Yau manifold $\Y$.
Depending on the specific form of the orientifold projection,
O3/O7- or O5/O9-orientifold planes are induced. These
negative tension objects are 
needed in brane-world scenarios with compact internal spaces 
to ensure cancellation of 
gravitational and electro-magnetic tadpoles.
Furthermore, the presence of localized sources 
requires a deviation from the standard Calabi-Yau compactifications
in that a non-trivial warp factor $e^{-2A}$ has to be included into the
Ansatz for the metric \cite{GKP,GP}
\beq
ds^2=e^{2A(y)} {\tilde g_{\mu \nu}}(x) dx^{\mu} dx^{\nu}+ e^{-2A(y)} 
     {g_{mn}}(y) dy^m dy^n\ ,
\label{metric}
\eeq
where $\tilde g_{\mu \nu}, \mu,\nu = 0,\ldots,3$ is a Minkowski metric 
and $ g_{mn}, m,n=1,\ldots,6$ is the metric on 
the Calabi-Yau manifold. 
However, in this paper we perform our analysis in the
unwarped Calabi-Yau manifold since in the large radius limit
the warp factor approaches one and the metrics 
of the two manifolds coincide  \cite{GKP,FP}. 
This in turn also implies that the metrics on the moduli space
of deformations agree and as a consequence the kinetic terms
in the low
energy effective actions are the same. The difference appears 
in the potential when some of the 
Calabi-Yau zero modes are rendered massive. 

In this paper we confine our attention to space-time filling
D3-branes and  
O3/O7 orientifold planes leaving a more general analysis
to a separate publication.
For this case the orientifold projection acting on the type IIB fields
is of the form \cite{Sen}--\cite{BH}
\bea 
  \mathcal{O} = (-1)^{F_L} \Omega_p \, \sigma^*, \label{o-projection}
\eea 
where $\Omega_p$ is the world-sheet parity and 
$F_L$ is the space-time fermion number
in the left-moving sector. 
$\sigma^*$ is the  pull-back of an isometric and holomorphic involution
acting on the Calabi-Yau manifold \cite{AAHV,BH}. 
This involution leaves the K\"ahler form $J$ of $\Y$ automatically  invariant
but can act non-trivially on the holomorphic three-form $\Omega$. 
O3/O7-orientifold planes are present when in addition
\beq\label{Omegatrans}
 \sigma^* \Omega = -\Omega  
\eeq
holds.\footnote{%
The case $\sigma^* \Omega = \Omega$ leads to 
O5/O9-planes and is analyzed in ref.\ \cite{GL} but plays no role in this paper.
Whenever $\sigma^* = id$ 
the theory has O9-planes and coincides with type I if one introduces D9-branes
to cancel tadpoles.}
Thus, we do not have to specify a particular Calabi-Yau manifold but merely
need to demand that it admits an isometric and holomorphic involution obeying
(\ref{Omegatrans}).
The analysis performed in this paper then 
holds for all Calabi-Yau manifolds with this property.

\subsubsection{The spectrum}\label{spectrum}

The massless spectrum of the effective low energy theory derived from such
an orientifold compactification is determined in ref.\ \cite{BH} and here
we need to briefly recall the result. Standard
Calabi-Yau compactifications of type IIB lead to an effective
$\cN=2$ supergravity in $D=4$ \cite{BGHL} which is further truncated to an 
$\cN=1$ supergravity by the orientifold projection.\footnote{Form
a supergravity point of view such truncations have been discussed
in ref.\ \cite{ADAF}.}
Focusing on the bosonic spectrum
one starts from the ten-dimensional
massless type IIB fields including the dilaton $\phi$, 
the metric $g$ and a two-form $B^{(2)}$ in the
NS-NS sector, and the axion $ l$, a second two-form 
$ C^{(2)}$ and a four-form $ C^{(4)}$ 
with a self-dual field strength in the R-R sector
which transform under (\ref{o-projection}) according to
\begin{equation}
\begin{array}{lcl}
  \mathcal{O}\  \phi &=& \ \sigma^*  \phi\ , \\
  \mathcal{O}\  g &=& \ \sigma^*  g\ , \\
  \mathcal{O}\  B^{(2)} &=& - \sigma^*  B^{(2)}\ ,
\end{array}
\hspace{2cm}
\begin{array}{lcl}
  \mathcal{O}\  l &=&  \ \sigma^*  l\ , \\
  \mathcal{O}\  C^{(2)} &=& - \sigma^*  C^{(2)}\ ,  \\
  \mathcal{O}\  C^{(4)} &=&  \ \sigma^*  C^{(4)}\ .
\end{array}
\label{fieldtransf}
\end{equation}
In the compactified theory these  ten-dimensional fields are 
expanded in terms of harmonic forms on $\Y$ and only the 
invariant states of 
the projection (\ref{o-projection}) together with (\ref{Omegatrans}) and
(\ref{fieldtransf}) are kept in the spectrum.
The harmonic forms are in one-to-one correspondence with the elements of 
the cohomology groups $H^{(p,q)}$ which 
split into two eigenspaces under the action of $\sigma^*$ 
\bea
  H^{(p,q)} = H^{(p,q)}_+ \oplus H^{(p,q)}_- \ .
\eea 
$H^{(p,q)}_+$ has  dimension $h_+^{(p,q)}$ and denotes
the $+1$ eigenspace of $\sigma^*$ while
$H^{(p,q)}_-$ has  dimension $h_-^{(p,q)}$ and denotes
the $-1$ eigenspace of $\sigma^*$. 
The Hodge $*$-operator commutes with $\sigma^*$ since $\sigma$ preserves the
orientation and the metric of the Calabi-Yau manifold and thus the Hodge
numbers obey $h^{(1,1)}_\pm=h^{(2,2)}_\pm$. Holomorphicity of $\sigma$ further implies
$h^{(3,0)}_\pm = h^{(0,3)}_\pm$ and $h^{(2,1)}_\pm = h^{(1,2)}_\pm$.
Combining these rules with the transformation properties (\ref{fieldtransf})
one can systematically determine the massless $D=4$ (bosonic) spectrum.
We will use the following basis for the spaces $H^{(p,q)}_{\pm}$:\\

\begin{table}[h] 
\begin{center}
\begin{tabular}{| c | c |c|c|c|c|} \hline
 \rule[-0.3cm]{0cm}{0.8cm} 
{\small $\omega_{a} \in H^{(1,1)}_- $}&
{\small ${\tilde \omega}^{a}\in H^{(2,2)}_- $}&
{\small $\chi_{\hat a} \in H^{(2,1)}_- $} &
{\small ${\bar \chi}_{\hab}\in H^{(1,2)}_- $} &
{\small $ \Ox \in H^{(3,0)}_-$}&
{\small $\Oxb \in H^{(0,3)}_- $}\\ \hline
\rule[-0.3cm]{0cm}{0.8cm} 
{\small $\omega_{\alpha}\in H^{(1,1)}_+ $}  & 
{\small ${\tilde \omega}^{\alpha}\in H^{(2,2)}_+ $}  &   
 {\small $\chi_{\hat \alpha} \in H^{(2,1)}_+ $} & 
{\small ${\bar \chi}_{\hat{\alpha}}\in H^{(1,2)}_+ $} &    &  \\
\hline
\end{tabular}
\caption{\label{table:basis}Basis for $H^{(p,q)}_{\pm}$}
\end{center}
\end{table}

The scalar fields arising from the metric are 
the deformations of the K\"ahler form $g_{k\bj}$ and 
the deformations of the complex structure 
which are proportional to $\delta{g}_{kj}$. 
Since the K\"ahler form  is left invariant by the orientifold
projection,  only 
$h_+^{(1,1)}$ K\"ahler deformations $v^\alpha$ survive and one has
\beq\label{transJ}
{g}_{k\bj}= -i v^{\alpha}(x)\, (\omega_{\alpha})_{k\bj} \ ,\quad
k,\bj=1,2,3\ ,\quad \alpha = 1,\ldots, h_+^{(1,1)}\ ,
\eeq 
where $\omega_\alpha$ denotes a basis of $H^{(1,1)}_+$.
Due to (\ref{Omegatrans}) the complex structure deformations of the metric
correspond to the elements in $H^{(1,2)}_-$ and one expands
\beq\label{cs}
\delta{g}_{kj} =  \frac{i}{||\Omega||^2}\, \bar z^{\hat a} 
(\bar \chi_{\hat a})_{k\ib\bl}\,
\Omega^{\ib\bl}{}_j \ , \quad \hat a=1,\ldots,h_-^{(1,2)}\ ,
\eeq
where $\bar\chi_{\hat a}$ denotes a basis of $H^{(1,2)}_-$ and we abbreviate
$||\Omega||^2\equiv \frac1{3!}\Omega_{ijk}\bar\Omega^{ijk}$.

Eqs.\ (\ref{fieldtransf}) further imply that the two-forms 
$B^{(2)}, C^{(2)}$
can only be expanded in terms of harmonic forms residing in $H^{(1,1)}_-$
while the four-form $C^{(4)}$ can be expanded in terms
 of harmonic forms in $H^{(1,1)}_+, H^{(2,2)}_+$ and  $H^{(3)}_+$.
One thus has
\bea\label{exp1}
  B^{(2)} &=& b^a(x)\, \omega_a\ ,\qquad C^{(2)}\ =\ c^a(x)\, \omega_a\ , 
\quad a=1,\ldots, h_-^{(1,1)}\ , \\
  C^{(4)} &=&  D_{(2)}^\alpha(x)\wedge \omega_\alpha+ V^{\hat \alpha}(x)\, \wedge \alpha_{\hat \alpha} + U_{\hat\alpha}(x)\wedge\beta^{\hat\alpha}+
 \rho_\alpha(x)\ \tilde \omega^\alpha\ ,\quad \hat \alpha = 1,\ldots,h_+^{(1,2)}\ ,\nonumber
\eea
where $b^a(x), c^a(x)$ and $\rho_\alpha(x)$  are space-time scalars,
$V^{\hat \alpha}(x)$ and $U_{\hat \alpha}(x)$ are  space-time one-forms 
and $D_{(2)}^\alpha(x)$ is a space-time two-form.
$\omega_a$ is a basis
of $H^{(1,1)}_-$, $\tilde\omega^\alpha$ is a basis
of $H^{(2,2)}_+$ which is dual to $\omega_\alpha$, and
$(\alpha_{\hat \alpha}, \beta^{\hat \alpha})$ is a real symplectic
basis of $H^{(3)}_+ = H^{(1,2)}_+ \oplus H^{(2,1)}_+$. 
Imposing the self-duality on the five-form field strength of $C^{(4)}$
eliminates half of the degrees of freedom in the expansion of 
$C^{(4)}$ leaving, for example,  only $\rho_\alpha$ and $V^{\hat \alpha}$
in the spectrum.

Let us summarize the low energy spectrum. Starting from the type IIB
massless fields and compactifying on a Calabi-Yau manifold
one keeps only fields that are invariant under the orientifold projections
(\ref{o-projection}), (\ref{Omegatrans}) and (\ref{fieldtransf}) and obtains
$h_+^{(2,1)}$ vectors  $V^{\hat \alpha}$,
$h_-^{(2,1)}$ complex scalars $z^{\hat a}$ 
parameterizing the deformations of the complex
structure,  $2 h^{(1,1)}_+$ scalars 
$(v^\alpha, \rho_\alpha) $ including the K\"ahler deformations $v^\alpha$,  
$2 h^{(1,1)}_-$ scalars $(b^a, c^a)$, the complex dilaton 
$\tau \equiv l + i e^{- \phi}$ and the space-time metric $g_{\mu\nu}$.
Including the fermions, these fields  assemble in an $\cN=1$ gravitational
multiplet, $h_+^{(2,1)}$ vector multiplets and 
$h_-^{(2,1)}+ h^{(1,1)}+1$ chiral multiplets (see Table 2.2). \\

\begin{table}[h] 
\begin{center}
\begin{tabular}{|l|c|c|} \hline 
 \rule[-0.3cm]{0cm}{0.8cm} 
 gravity multiplet&1&$g_{\mu \nu} $ \\ \hline
 \rule[-0.3cm]{0cm}{0.8cm} 
 vector multiplets&   $h_+^{(2,1)}$&  $V^{\hat \alpha} $\\ \hline
 \rule[-0.3cm]{0cm}{0.8cm} 
 \multirow{4}{30mm}[-3.5mm]{chiral multiplets} &   $h_-^{(2,1)}$& $z^{\hat a} $ \\ \cline{2-3}
 \rule[-0.3cm]{0cm}{0.8cm} 
 &     $h^{(1,1)}_+$& $( v^\alpha, \rho_\alpha )$ \\ \cline{2-3}
 \rule[-0.3cm]{0cm}{0.8cm} 
 &  $ h^{(1,1)}_-$ &$( b^a, c^a)$ \\ \cline{2-3}
 \rule[-0.3cm]{0cm}{0.8cm} 
   & 1 & $(\phi,l)$ \\ \hline
\end{tabular}
\caption{$\cN =1$ spectrum of moduli.}
\end{center}
\end{table}

\subsubsection{The effective action without background fluxes}
The low energy effective action of the massless modes is
determined by a
Kaluza-Klein compactification of the ten-dimensional type IIB supergravity.
This reduction is carried out in detail in ref.\ \cite{GL} 
and briefly summarized in appendix \ref{KKreduction}.
 The resulting four dimensional supergravity theory  can be written in the 
standard $\cN=1$ form 
\begin{eqnarray}\label{N=1action_WB}
  S^{\rm E}_{\rm Bulk} &=& \int_{\mathbb{M}_{3,1}} - \frac{1}{2}R * \mathbf{1} -
  \hat K_{I \bar J}\, dM^I \wedge * d\bar M^J  -
  e^{\hK} \left(\hat K^{I \bar J} D_I \hat W \bar D_{\bar J} \hat{\bar W}-3|\hat W|^2 \right)
\nonumber \\
  && \qquad - \frac{1}{2}(\text{Re}\; f_{\hat \alpha \hat \beta})\ 
  \FA^{\hat \alpha} \wedge * \FA^{\hat \beta}  
  - \frac{1}{2}(\text{Im}\; f_{\hat  \alpha \hat \beta})\ 
  \FA^{\hat \alpha} \wedge \FA^{\hat \beta} \ , 
\end{eqnarray}
where $M^I, I = 1, \ldots, h_-^{(2,1)}+ h^{(1,1)}+1$ 
collectively denotes all scalar fields in chiral multiplets and 
$\FA^{\hat \alpha}=dV^{\hat \alpha}$ is the field strength of the 
$h_+^{(2,1)}$ (Abelian) vector multiplets.
$\hat K_{I \bar J}= \partial_I \bar\partial_{\bar J}\hat K(M^I, \bar M^I)$ 
 is the 
K\"ahler metric (with K\"ahler potential $\hat K$)
on the moduli space of the compactification while $f(M^I)$ and  $\hat W(M^I)$
are the holomorphic gauge kinetic function and the holomorphic superpotential,
respectively.\footnote{We use the notation $\hat K, \hat W$ in order to
distinguish from $K,W$ used later on when also matter fields arising from 
D-branes are included.}
The action given 
in terms of the scalar fields arising in the expansions 
(\ref{transJ})--(\ref{exp1}) is discussed
in appendix \ref{KKreduction}.
To obtain the standard form (\ref{N=1action_WB}),
a (complicated) field redefinition is necessary.
One finds that the K\"ahler structure of the moduli space 
is manifest in the complex coordinates 
$M^I= (\tau,G^a,T_\alpha,z^{\hat a})$ defined as
\begin{eqnarray}\label{tau}
  \tau &=& l+i e^{- \phi}\ , \qquad
  G^a = c^a -\tau b^a\ ,\\
  T_\alpha &=& \frac{3i}{2} \rho_\alpha 
  + \frac{3}{4}\cK_{\alpha} - \frac{3i}{4(\tau-\bar \tau)}\,  \cK_{\alpha b c}G^b (G- \bar G)^c  \  , \nonumber
\end{eqnarray}
where  we abbreviate
$\cK_{\alpha} \equiv \cK_{\alpha \beta \gamma} v^\beta v^\gamma$.
$\cK_{\alpha \beta \gamma}$ and $\cK_{a b \gamma}$ are (constant) intersection
numbers  defined as 
\bea\label{int-num}
\cK_{\alpha \beta \gamma}=\int_{\Y} \omega_\alpha \wedge \omega_\beta \wedge
                                         \omega_\gamma\ , \qquad
  \cK_{a b \gamma}=\int_{\Y} \omega_a \wedge \omega_b \wedge
                                         \omega_\gamma\ ,
\eea
which are the only non-vanishing intersection numbers after the orientifold projection
\cite{ADAF,GL}. 
In terms of the coordinates defined in (\ref{tau}), the 
K\"ahler potential is given by 
\begin{eqnarray}
 \hK=-\text{ln}\Big[-i\int\Omega(z) \wedge \bar \Omega(\bar z) \Big] 
  -\text{ln}\big[-i(\tau - \bar \tau)\big]
  - 2 \text{ln}\Big[\frac16\cK(\tau,T,G)\Big]\ ,
  \label{kaehlerpot-O7-1}
\end{eqnarray}
where 
$\cK \equiv \cK_{\alpha \beta \gamma} v^\alpha v^\beta v^\gamma = 6\, \text{Vol}(\Y)$ 
is related to the 
volume of the Calabi-Yau manifold. $\cK$ should be understood
as a function of the K\"ahler coordinates $(\tau,T,G)$ 
which enter by solving (\ref{tau}) for $v^\alpha$ in terms of $(\tau,T,G)$.
Unfortunately this solution cannot be given explicitly and therefore $\cK$ is known
only implicitly via $v^\alpha(\tau,T,G)$.\footnote{This 
is in complete analogy
to the situation encountered in compactifications
of M-theory on Calabi-Yau fourfolds studied in \cite{HL}.}
However, for one overall K\"ahler metric modulus $v$ parameterizing the volume
(i.e. for $h^{(1,1)}_+=1$, $T_{\alpha}\equiv T$), keeping 
all $h^{(1,1)}_-$ moduli, eq.\ \eqref{tau} can be solved for
$v$ and one finds
\beq
  -2 \ln\; \cK = -3 \ln \frac{2}{3}\left[ T + \bar T
  + \frac{3i}{4(\tau - \bar \tau)} \cK_{1 a b} (G-\bar G)^a (G-\bar G)^b \right]\ .
\eeq
which has the standard no-scale structure.

The first two terms in (\ref{kaehlerpot-O7-1}) are the standard
K\"ahler potentials for the complex structure deformations
and the dilaton, respectively. 
$\cK$ also depends on $\tau$ and therefore the metric mixes $\tau$
with $T_\alpha$ and $G^a$. It is block diagonal in the 
complex structure deformations which do not mix with the other scalars.
Thus, the moduli space has the form
\beq \label{modulispace}
\cM = \cM^{h_-^{(1,2)}}_{\rm cs}\times\, \cM^{h^{(1,1)_{\vphantom{+}}}+1}_{\rm k}\ ,
\eeq
where each factor is a K\"ahler manifold.\footnote{In the next section we will see
that including matter fields from the D-branes also mixes the complex structure
deformations non-trivially with all the other moduli and the product structure is lost.}

For completeness let us also give the
gauge-kinetic coupling functions $f_{\hat \alpha \hat \beta}$ which
only depend on $z^{\hat a}$ but not on any of the other moduli.
They are given by
\beq\label{last}
  f_{\hat \alpha \hat \beta}= 
       -\frac{i}{2}\, \cF_{\hat \alpha \hat \beta}\ ,
\eeq
where $\cF_{\hat \alpha \hat \beta}$ is a holomorphic function of 
the complex structure deformations $z^{\hat a}$.
It is computed as a second derivative
from the full $\cN=2$ prepotential $\cF^{\cN=2}(z)$ of all
complex structure deformation via \cite{ADAF}
\beq
  \cF_{\hat \alpha \hat \beta} 
= \partial_{z^{\hat\alpha}} \partial_{z^{\hat \beta}}\cF^{\cN=2}(z^{\hat\alpha},z^{\hat a})|_{z^{\hat\alpha}=0}\ .
\eeq
Since we do not need these couplings in this paper we refer the reader to \cite{GL}
for further details.

Finally, the superpotential vanishes as long as no background fluxes are 
turned on and $\cN=1$ supersymmetry remains unbroken. 
Let us now turn to the situation when non-trivial background fluxes are
present.

\subsubsection{The effective action with non-trivial background fluxes}
\label{subsubsec:fluxes}

We are still in the process of preparing the ground for the next section where
we add space-time filling D3-branes into the picture. For consistency, this requires 
the presence of negative tension objects such as orientifolds and non-trivial
five-form flux. Such configurations break $\cN=2$ supersymmetry of the
original Calabi-Yau compactification to $\cN=1$.
In addition, one can also
turn on three-form flux which induces a superpotential. This
leads to the stabilization of part of the moduli and allows
for the possibility of spontaneous $\cN=1$ supersymmetry breaking.

Let us first discuss the five-form flux.
The self-dual field strength of the four-form $C^{(4)}$ is defined (in $D=10$) as
${{\tilde F}}^{(5)}= {F}^{(5)} -\frac{1}{2}C^{(2)} \wedge H^{(3)} 
+\frac{1}{2}  B^{(2)} \wedge  F^{(3)}$ where ${F}^{(5)}=dC^{(4)}$,
$H^{(3)}= dB^{(2)}$
and $F^{(3)} = dC^{(2)}$. Since
D3-branes and O3-planes are 4D Poincar\'e invariant sources,  
the 5-form flux should respect this symmetry. Thus the
only possible flux that can be turned on is of the form 
\beq
{\tilde F}^{(5)}=dC^{(4)} 
\, ,  \qquad 
C^{(4)}=\alpha(y) \, dx^0\wedge dx^1 \wedge dx^2 \wedge dx^3\ .
\label{F5}
\eeq

In addition to the five-form flux one can also turn on
three-form fluxes $H^{(3)}$ and $F^{(3)}$ on the Calabi-Yau manifold $\Y$.  
The Bianchi identity  combined with the equations of motion
imply that both fluxes have to be harmonic three-forms and therefore
can be parameterized in terms of the third cohomology group $H^3(\Y)$.
For the orientifold compactifications under consideration
they also have to respect the orientifold projection and remain invariant
under the action of ${\cal O}$. This implies that they can only take values in
$H^3_-(\Y)$.

The five-form flux $\alpha$ is further constrained by the Bianchi identity
and the trace of the Einstein equations.
Taking the difference between these two equations one obtains \cite{GKP}
\beq
{\nabla}^2(e^{4A}-\alpha)= e^{-6A} \left|\partial(e^{4A} -\alpha)\right|^2 + \frac{e^{2A} e^{\phi}}{6} 
\left|*_6 G^{(3)} - i G^{(3)}\right|^2\ ,
\label{e4A-alpha}
\eeq
where 
\beq\label{Gdef}
G^{(3)}\equiv F^{(3)} - \tau H^{(3)}\ ,
\eeq 
and $*_6$ is the six-dimensional 
Hodge $*$-operator on $\Y$.\footnote{Note that for 
the Bianchi identity of the five-form flux and the
Einstein equations there should be source terms. 
They cancel in the difference
$\nabla^2 (e^{4A} -\alpha)$ 
in the RHS of (\ref{e4A-alpha}) for D3-branes and O3-planes \cite{GKP}.}
On a compact space,  the LHS of (\ref{e4A-alpha})
integrates to zero, while the RHS 
is non-negative. This implies  a relation between 
$\alpha(y)$ and the warp factor $e^{-2A(y)}$
\beq
e^{4A}=\alpha
\label{e4A=alpha}\ 
\eeq
and 
\beq\label{ISD}
*_6 G^{(3)}=iG^{(3)}\ .
\eeq 
Fluxes which obey this condition are called imaginary 
self-dual (ISD). The equation of motion for the dilaton 
is solved by constant $\tau$ and as a consequence 
$G^{(3)}$ is harmonic since
both $F^{(3)}$ and  $H^{(3)}$ are harmonic three forms.
The Hodge operator $*_6$ acts on such forms according to
\bea
*_6\Ox&=&-i \Ox, \ \ \  \ \ \ *_6 \cha = \ i \cha, \nn \\
*_6\Oxb&=& \ i \Oxb, \ \ \  \ \ \ \   *_6 \chab = -i \chab \ ,
\label{ISDIASD}
\eea
where $\Omega$ is the $(3,0)$ form while $\cha$ are $(2,1)$ forms introduced in  
Table \ref{table:basis}.
This implies that the $G^{(3)}$ which obeys (\ref{ISD}) is a sum 
of $(2,1)$ and $(0,3)$ forms only, and for these 
cases  a consistent supergravity background on a warped compact Calabi-Yau orientifold exits.\footnote{Other sources such as anti-D3-branes or
$\bar O$-planes do not cancel in the difference (\ref{e4A-alpha}).
Anti-D3-branes give a positive contribution to the RHS,
so they do not help evade the no-go result, but $\bar O_+$-planes
(those with negative tension and positive charge) give a negative
contribution \cite{Buchel}  (we thank E.Dudas, A.Frey and E.Kiritsis for
discussions on this point).
In this paper we consider only O3-planes and D3-branes,
leaving this possibility for future investigation.}

The supersymmetry transformations in this background were analysed
in refs.\ \cite{GP,Gubser}. It was shown that a primitive $(2,1)$ piece of
the three-form flux preserves the $\cN=1$ supersymmetry while
any other three-form flux breaks it.
Let us already mention that including D-branes does not change
this conclusion as can be seen from eq.\ (\ref{susy}).  
This implies that when we turn on (2,1) three-form flux, 
the  CY-orientifold including D3-branes  is 
$\cN=1$ supersymmetric.\footnote{The primitivity condition  
on the 3-form flux $G^{(3)} \wedge J=0$ 
is automatically satisfied on a {\CY}.}    
On the other hand $(0,3)$ flux does break supersymmetry 
spontaneously with vanishing cosmological constant \cite{GKP}.
This fact is best seen from the effective action including
background fluxes.

The Kaluza-Klein reduction briefly reviewed in the previous
section can also be performed when the three-form flux $G^{(3)}$
is non-vanishing. The K\"ahler potential and the gauge kinetic function
of the previous section are unchanged\footnote{This again assumes 
a small warp factor. Including the warp factor consistently  is beyond
the scope of this paper. The issue 
has been discussed in refs.\ \cite{DWG,deAlwis}. 
The point is that the harmonic analysis
performed in the previous subsection 
has to be reconsidered and appropriately adjusted to 
warped compactifications on
conformal Calabi-Yau manifolds \cite{GSS}.} but a non-trivial
superpotential is induced. One finds\footnote{%
This superpotential was first suggested in ref.\ \cite{GVW};
for orientifold backgrounds it was rederived in \cite{GL}.}
\beq
\hat{W}= \int \Omega \wedge G^{(3)}\ ,
\label{hatW}
\eeq
which depends on the $h^{(1,2)}_-$ 
complex structure moduli $z^{\hat a}$ through $\Ox$ and on $\tau$ 
through the definition of $\Gt$. It vanishes for $(2,1)$ flux 
(and also for $(3,0)$ and $(1,2)$ flux) but
is non-zero for (0,3)-flux. 

The K\"ahler covariant derivatives 
$D_I \hat W = \partial_I \hat W + \hat W \partial_I \hat K$
are the order parameters for spontaneous supersymmetry
breaking. For the case at hand these derivatives are evaluated in 
(\ref{kcov1}) and one sees that they vanish if 
\beq
\hat{W}=0\ , \qquad 
\G \equiv \int {\bar \Omega} \wedge G^{(3)} =0\ , 
\qquad \CHI_{\ha}\equiv \int \chi_{\ha} \wedge G^{(3)}=0\ ,
\label{susyfromW}
\eeq
hold. $\hat W$ is 
non-vanishing for (0,3)  flux,  $\G$ is non-vanishing for (3,0)
flux and $\CHI_{\ha}$ is non-vanishing for (1,2) flux 
or in other words $\hat W,\G$ and $\CHI_{\ha}$ are integral
representations of (0,3), (3,0) and (1,2) flux, respectively.
Hence, supersymmetry is unbroken whenever
the (0,3), (3,0) and (1,2) pieces of $G^{(3)}$ vanish,
while the (2,1) piece can be arbitrary.
In this case also $\hat W$ is zero implying the vanishing
of the cosmological constant or in other words the existence of
a Minkowskian supersymmetric ground state in full agreement with
the result of ref.\ \cite{GP}. 
Conversely, any (0,3), (3,0) and (1,2) pieces of $G^{(3)}$
break supersymmetry spontaneously since they lead a non-zero
$D_I \hat W$. 

In order to determine the cosmological constant
for these cases, we need to evaluate the scalar potential.
It  can be computed from (\ref{N=1action_WB})
using (\ref{kaehlerpot-O7-1}), (\ref{hatW}),  (\ref{Kinvers}) and 
(\ref{kcov1}), or directly from a Kaluza-Klein reduction. 
Both computations yield 
\bea
\hat{V}&=&\frac{ 18 i e^{\hphi}}{\cK^2 \int \Ox \wedge \Oxb } \left(\int \Ox \wedge \Gtb \int \Oxb \wedge \Gt + \Gw^{\ha \hbb}
\int \chi_{\ha} \wedge \Gt \int \bar{\chi}_{\hbb} \wedge \Gtb \right)\nn \\
&=& e^{\hK} \left( \,|\G|^2\,+\,\Gw^{\ha \hbb}\, \CHI_{\ha} \,\CHI_{\hbb} \right)\ ,
\label{Vhat}
\eea
where $\Gw^{\ha \hbb}$ is defined in \eqref{metrics}.
%
%
We see that the potential does not depend on $\hat W$,
but only on the two other integrals $\G$ and $\CHI_{\ha}$.
Thus,
for (0,3) flux
the potential vanishes identically which, in fact, 
is an example of 
``no-scale'' supersymmetry breaking \cite{NS, GKP}.   
The (3,0) and (1,2)-pieces on the other hand,  
contribute  positive semi-definite terms 
to the potential and therefore correspond to 
spontaneous supersymmetry breaking in a Minkowski
or de Sitter background. Due to the overall dilaton and volume
dependence generically a `run-away' solution will force
the potential to zero for zero string coupling or infinite volume.
This instabilty is just another manifestation of the fact
that within our setup,  eq.\ (\ref{e4A-alpha}) does not
allow imaginary anti-self dual (IASD) 3-form fluxes, i.e.\ 
the (1,2) and (3,0) pieces of $G^{(3)}$. 

The potential (\ref{Vhat}) generated by three-form fluxes
stabilizes the dilaton and the complex structure deformations, but 
leaves the K\"ahler moduli unfixed. 
Modifying the setup by other localized sources and/or
including other non-perturbative effects
is beyond the scope of this paper. However, it is still
interesting to study the structure of soft supersymmetry breaking 
which is induced by IASD fluxes under the assumption
that the present configuration is rendered stable by other effects.
In other words we 
can continue under the assumption
that further terms in the superpotential are generated
which stabilize the moduli but otherwise do not 
interact with the matter fields on the D-branes
which we are going to discuss in the next section.
In terms of eq.\ (\ref{e4A-alpha}), this corresponds
to its local solution without taking into account
the global properties. As we will see in section~3, this is enough
to see the structure of the soft terms.

\subsection{The Brane: D3-branes coupled to Calabi-Yau orientifolds}
\label{subsec:D3action}

In this section we derive the low energy effective action of a stack of 
$N$ space-time filling D3-branes in type IIB string theory 
compactified on a Calabi-Yau orientifold as described in the previous section. Our starting point for the bosonic terms 
is the non-Abelian Dirac-Born-Infeld and Chern-Simons action as 
proposed in ref.\ \cite{Myers}. 
These action functionals are expanded to fourth order in the fields including the fluctutations of the brane position in the internal space $Y$. In the effective action, the fluctuations 
give rise to scalar fields $\phi$ 
charged under a non-Abelian gauge group, which in our case will be $U(N)$. The supersymmetric extension of this action leads to
appropriate fermionic fields which combine with the $\phi$
to form $\cN=1$ chiral superfields. They can be viewed as the 
charged matter multiplets of the theory coupling to
the bulk moduli introduced in the previous section.
The purpose of this section is to derive the effective action for the matter fields $\phi$ (and their superpartners)
and their couplings to the moduli.
Apart from the supersymmetric terms we also compute
moduli-dependent masses and trilinear 
couplings which arise as a consequence of spontaneous supersymmetry
breaking.

\subsubsection{Bosonic action}\label{subsubsec:D3bosons}

The bosonic part of the action of a single D$p$-brane is captured by the Dirac-Born-Infeld action, which reads in string frame
\begin{equation}
   S_{\text{DBI}}^{\text{sf}}=-\mu_p\int_\WV d^{p+1}\xi\:e^{-\phi}
         \sqrt{-\det\left(\Em^*E_{\mu\nu}+\ell \FD_{\mu\nu}\right)}\ ,
\end{equation}
and the topological Chern-Simons action
\begin{equation} \label{eq:GeneralChernSimon}
   S_{\text{CS}}=\mu_p\int_\WV\Em^*\Big(\sum_q C^{(q)}e^B\Big)
\,e^{\ell\FD}\  ,
\end{equation}
where $\ell=2\pi\alpha'$. The absolute value of the RR-charge $\mu_p$ of the brane is equal to the brane tension for BPS branes. The integrals are taken over the $p+1$-dimensional world-volume $\WV$ of the D$p$-brane, which is embedded in the ten dimensional space-time manifold $M$ via the map $\Em:\WV\hookrightarrow M$. 
For simplicity the combination of the metric $g$ and the $B$-field  
\begin{equation}
   E_{\mu\nu}=g_{\mu\nu}+B_{\mu\nu}
\end{equation}
is used. In the brane action the bulk fields $E$ and the bulk RR-fields 
$C^{(q)}$ are pulled back with $\Em^*$ to the world-volume $\WV$ of the brane. The dynamics of the D$p$-brane is encoded in the pull-back of $E$.
 
The Dirac-Born-Infeld action contains a $U(1)$ field strength $\FD$, which describes the $U(1)$ gauge theory of the endpoints of open-strings attached to the brane to all orders in $\alpha' \FD$
\cite{Leigh:jq}. 
To leading order, the gauge theory reduces to a $U(1)$ Yang-Mills theory on the world-volume $\WV$ of the brane.  
  
Since D$p$-branes carry RR-charges \cite{JP}, they couple as extended objects to appropriate RR-forms of the bulk, namely the $p+1$ dimensional world-volume couples naturally to the RR-form $C^{(p+1)}$. 
Moreover, generically D-branes contain lower dimensional D-brane charges, and hence interact also with lower degree RR-forms 
\cite{Douglas:1995bn}. 
All these couplings to the bulk are implemented in the Chern-Simons action in a way compatible with T-duality. In type IIB string theory we only have even degree RR-forms $C^{(q)}$, and as a
consequence we only can have D$p$-branes for odd $p$, in particular D3-branes. Furthermore it implies that in type IIB the sum in \eqref{eq:GeneralChernSimon} runs only over even forms. Note that $C^{(0)}$ is the axion 
which we called $l$ in the previous section. 

In order to describe a stack of D$p$-branes the above action has to be generalized. The endpoints of open strings are now labeled by the brane they are attached to. Therefore, the modified world-volume action must be a non-Abelian gauge theory of these labels, which are called Chan-Paton factors.  In type IIB compactifications with orientifolds, the possible gauge groups turn out to be $U(N)$, $SO(N)$ and $Sp(N)$. For branes which coincide with orientifold planes the gauge groups are either $SO(N)$ or $Sp(N)$ depending on the type of orientifold plane\cite{BH}. Branes that are not invariant under the orientifold projection have $U(N)$ as a gauge group. 
We limit our analysis to the latter case and from now on
consider a stack of $N$ space-time filling D3-branes.
 
For the non-Abelian Dirac-Born-Infeld action 
we use the form\cite{Myers}
\begin{align}
   S_{\text{DBI}}^{\text{sf}}= 
     -\mu_3\int_\WV d^{4}\xi\:\tr e^{-\phi} 
     \sqrt{ -\det \Big( \Em^* \big(E_{\mu\nu}+E_{\mu n}(Q^{-1}-\delta)^{nm}E_{m\nu}\big)
     +\ell\FD_{\mu\nu}\Big)\det Q^n_m}\ , \label{eq:DBIaction}
\end{align}
where
\begin{equation}
   Q^n_m=\delta^n_m+i \ell \com{\phi^n}{\phi^k} E_{km} \ , \quad
n,m=1,\ldots,6\ .
\label{eq:Qterm} 
\end{equation}
Now the function $\Em:\WV\hookrightarrow M$ describes the embedding of the stack of branes in the space-time manifold $M$.
The six $\phi^n$ parameterize the fluctuations of the branes 
and are in the adjoint representation of $U(N)$.
For the Chern-Simons action we use \cite{Myers} 
\begin{equation}
   \mathcal{S}_{\text{CS}}=\mu_3 \int_\WV\tr 
     \Big(\Em^*\big(e^{i\ell \ins{\phi}\ins{\phi}}\sum_{q\, {\rm even}}
  C^{(q)} e^B\big)\,
     e^{\ell\FD} \Big)\ . \label{eq:ChernSimon}
\end{equation}
$\ins{\phi}$ denotes the interior multiplication of a form with $\phi^n$, which yields for a local $q$-form 
\begin{equation} \label{eq:intmul}
   \ins{\phi}C^{(q)}=\frac{1}{q!}\sum_{k=1}^q (-1)^{k+1}\phi^n 
      C^{(q)}_{\nu_1\ldots\nu_{k-1}\:n\:\nu_{k+1}\ldots\nu_q}
       d x^{\nu_1}\wedge\ldots\widehat{ d x^{\nu_k}}
      \ldots\wedge d x^{\nu_q}  \ ,
\end{equation}
where the differential with the hat $\widehat{\quad}$ is omitted. 
Note  that in equations \eqref{eq:DBIaction} and \eqref{eq:ChernSimon} the symmetrized average \cite{Tseytlin} of the trace has to be taken with respect to the non-Abelian expressions $\FD_{\mu\nu}$, $\D_\mu\phi^m$ and $\com{\phi^m}{\phi^n}$. 

Already in the Abelian case one expects additional corrections in $\alpha'$ involving derivative terms beyond second order. However, the Abelian Dirac-Born-Infeld action is expected to capture all $\alpha'$ corrections in $\FD$ for ``slowly-varying'' $\FD$ 
(i.e.\ all derivative-independent terms in $\FD$). In the non-Abelian case the distinction between the field strength and its covariant derivative is ambiguous since $[\D_{\mu},\D_{\nu}]\FD_{\lambda \rho}=[\FD_{\mu \nu},\FD_{\lambda \rho}]$. The
symmetrized trace proposal of ref.\
\cite{Tseytlin} treats the matrices $\FD$ (as well as $\D_\mu\phi^m$ and $\com{\phi^m}{\phi^n}$) as if they were commuting,
leaving out all commutators among these. This proposal was shown to be reliable only up to fourth order in $\FD$ \cite{HT}, 
but this is enough for our purpose.

In order to expand  the Dirac-Born-Infeld action \eqref{eq:DBIaction} we first have to expand the square root of the determinant using 
 the standard formula
\begin{equation}
   \sqrt{\det\left(\id+M\right)}\ =\ 
1+\frac{1}{2}\tr M-\frac{1}{4}\tr M^2+
     \frac{1}{8}\left(\tr M\right)^2+\ldots \  .
\end{equation}
Second, we need to evaluate 
the pull-back of the metric which is carefully derived in appendix
\ref{pullback}. In the Abelian case we obtain 
from eq.\ \eqref{eq:gpbcomp} 
for the warped metric 
\begin{equation}
   \varphi^*(g)_{\mu\nu}=e^{2A(y_0)} \tilde g_{\mu\nu}+e^{-2A(y_0)}\ell^2\ g_{mn}
     \D_\mu\phi^m\D_\nu\phi^n+
     e^{2A(y_0)}\ell^2 \tilde g_{\mu\tau}\Riem{n}{\tau}{\nu}{m}\phi^n\phi^m \ ,
\end{equation}
where $y_0$ denotes the locus of the D3-branes in the internal 
space.\footnote{We choose $y_0$ not to be a fixed point of the orientifold involution so as to get a $U(N)$ gauge theory on the world-volume of the branes.}
$\D_\mu$ is the covariant derivative with respect to the gauge group, and moreover it contains a connection of the normal bundle
of the D3-brane \cite{HM}. The latter connection is, however, trivial in the limit of vanishing warp factor due to the product ansatz of the ten dimensional metric \eqref{metric}. 

The non-Abelian nature of the $\phi$ are taken into account
by using a non-Abelian Taylor expansion 
of the background fields
\cite{Myers,Douglas:1997ch,Garousi:1998fg}. 
On a generic background field $T$ this expansion yields
\begin{equation} \label{eq:nonabelianTaylor}
   T=\exp\left[\ell\phi^n\partial_n\right]\:
       \left.\varphi^*T\right|_{y_0} 
     =\sum_{k=0}^\infty\frac{\ell^k}{k!}\,\phi^{n_1}\cdot\ldots\cdot\phi^{n_k}
       \partial_{n_1}\ldots\partial_{n_k}
       \left(\left.\varphi^*T\right|_{y_0}\right)\  .
\end{equation}
Applying this non-Abelian Taylor expansion to the determinant of \eqref{eq:Qterm} we obtain
\begin{equation}
   \sqrt{\det Q^i_j}\
=\ 1+\frac{i\ell^2}{2}\com{\phi^m}{\phi^n}\phi^k\partial_kB_{nm}
     +\frac{\ell^2}{4} g_{mn}g_{op}\com{\phi^o}{\phi^m}\com{\phi^n}{\phi^p}+\ldots\  ,
\end{equation}
where $\ldots$ denotes terms which vanish after taking the trace in the Lagrangian.  Assembling all terms together we arrive at
the following action in the 4D-Einstein frame \cite{MG}
\begin{equation}  \label{eq:DBIlowone}
\begin{split}
   S_{\mathrm{DBI}}^{\mathrm{E}}
     =-\mu_3 \int_\WV d^4\xi&\sqrt{-g_4}\
      \tr\left(\frac{36\:e^{4A}}{\Kw^2}(1+\ell^2\Riem{n}{\tau}{\tau}{m}\phi^n\phi^m)
       +\frac{\ell^2}{4}e^{-\phi} F^{\mu\nu}F_{\nu\mu} \right. \\
      &+\frac{3\ell^2 }{\Kw}\, g_{mn}
        \D_\mu\phi^m \D^\mu\phi^n 
       +\frac{9\ell^2e^{4A}}{\Kw^2}e^{\phi}(G^{(3)}-\bar G^{(3)})_{lmn} 
        {\phi^l}{\phi^m}{\phi^n} \\
      &+\frac{9\:\ell^2}{\Kw^2}e^{\hat\phi}\left.  
       g_{qp}g_{mn} 
             \com{\phi^q}{\phi^m}\com{\phi^{n}}{\phi^{p}}
             \vphantom{\frac{e^{4A}}{\Kw^2}}\right) \ ,
\end{split}
\end{equation}
where we used \eqref{Gdef}.  Note that  
the last two terms vanish in the Abelian limit.

$g_4$ is the determinant of the 4D-Einstein frame metric, which is related to
the metric ${\tilde g}_{\mu\nu}$ defined in (\ref{metric}) by \cite{DWG}
\beq
g_{\mu \nu}=\dfrac{1}{6}\, \Kw{\tilde g}_{\mu \nu}\ , \qquad
\Kw=6 \int_Y d^6 y \sqrt{det g_{mn}} e^{-4A} \ .
\eeq
The factor of $e^{-4A}$ results from the reduction of
the ten-dimensional curvature scalar of the warped metric.
In the reduction of the brane action we include the warp factor
taking the large volume  limit only when we combine it
with the bulk action.

Since the $\phi^m$ are scalar components of chiral superfields
they have to combine into complex variables. 
Therefore we  have to rewrite the action (\ref{eq:DBIlowone})
in terms of complex fields or in other words we have to find
a complex structure compatible with $\cN=1$ supersymmetry.
{}From the action \eqref{eq:DBIlowone} we see that
the $\sigma$-model metric of the $\phi^m$
coincides with the Calabi-Yau metric $g_{mn}$ and thus a natural
guess is to choose the  complex
structure $J$
of $\Y$ also as the complex structure of the low
energy effective action.\footnote{By abuse of notation 
we use the same 
symbol $J$ as we used for the  K\"ahler form in section 2.1.}
For fixed complex structure we just rewrite all equations in terms of complex indices, i.e. we choose a basis which is compatible with the complex structure $J$. With respect to this basis $J$ takes block diagonal form
\begin{equation}
   J=\begin{pmatrix} +i\id & \\ & -i\id \end{pmatrix}  .
\end{equation}
Including the complex structure deformations to lowest order
we have to perturb $J$ according to \cite{GSWVolume2}
\begin{equation}
   \tilde{J}= J+\delta J = 
      \begin{pmatrix} 
          +i\id & z^{\hat a}\chi_{\hat a} \\ 
          \bar z^{\hat a}\bar\chi_{\hat a} & -i\id  
      \end{pmatrix}\  ,
\end{equation}
where $\chi_{\hat a}$ is an element of 
$H^{(0,1)}(Y,T^{(1,0)})$ related to the basis of $H^{(2,1)}_-$
defined in Table \ref{table:basis} via
\beq
(\chi_{\hat a})_{\bj}^i
   = \frac{1}{\norm{\Omega}^2}\, \bar\Omega^{ilk} 
      (\chi_{\hat a})_{lk\bar j} \ .
\eeq

As we perturb the complex structure $J$ to $\tilde J$, the eigenvectors of $J$ are also modified. To first order the perturbed eigenvectors read
\begin{equation}
   \begin{pmatrix} \phi \\ 0 \end{pmatrix} \rightarrow 
      \begin{pmatrix} \phi \\ -\frac{i}{2}\bar z^{\hat a}\bar\chi_{\hat a}\phi \end{pmatrix}\ , \qquad
   \begin{pmatrix} 0 \\ \bar\phi \end{pmatrix} \rightarrow
      \begin{pmatrix} \frac{i}{2}z^{\hat a}\chi_{\hat a} \bar\phi \\ \bar\phi \end{pmatrix} \ ,
\end{equation}
with $\phi$ a vector of $T^{(1,0)}$ and $\bar\phi$ a vector of $T^{(0,1)}$ with respect to the fixed complex structure $J$. Furthermore $\chi_{\hat a}$ maps a tangent vector of type $(0,1)$ to a tangent vector of type $(1,0)$, and $\bar\chi_{\hat a}$ vice versa. Hence the complex structure deformations act (up to first order) on the total vector $\phi^n$ in component notation as 
\begin{equation}
 \label{eq:Vectpertcomp}
   \phi^i\rightarrow\phi^i+\frac{i}{2}z^{\hat a}(\chi_{\hat a})_{\bar l}^i\,\bar\phi^{\bar l}\ ,
\qquad
   \bar\phi^{\bj}\rightarrow\bar\phi^{\bj}
      -\frac{i}{2}{\bar z}^{\hat a}(\bar\chi_{\hat a})_l^{\bj}\,\phi^l  
\ .
\end{equation}
Thus in the kinetic term in the action (\ref{eq:DBIlowone})
we have to replace 
\begin{equation}\label{gdefo}
   g_{mn}\rightarrow g_{i\bj}+g_{\bj i}+
        \delta g_{ij}(\bar z^a)+\delta g_{\bar\imath\bj}(z^a)\  ,
\end{equation}
 and simultaneously
\begin{equation} \label{eq:Vectpertderiv}
\begin{split}
   \D_\mu\phi^i &\rightarrow \cD_\mu\phi^i+\frac{i}{2} z^{\hat a}
       (\chi_{\hat a})_{\bar l}^i\,\D_\mu\bar\phi^{\bar l}\ , \\
   \D_\mu\bar\phi^{\bj}&\rightarrow\cD_\mu\bar\phi^{\bj}
      -\frac{i}{2}\bar z^{\hat a}(\bar\chi_{\hat a})_l^{\bj}\,\D_\mu\phi^l\  ,
\end{split}
\end{equation}   
where $\delta g_{ij}$ and $\delta g_{\bar\imath\bj}$ 
are defined in \eqref{cs}, and the covariant derivatives read
\begin{equation} \label{eq:VectcovderivTM}
\begin{split} 
   \cD_\mu\phi^i &= \D_\mu\phi^i+\frac{i}{2}\partial_\mu z^{\hat a}
       (\chi_{\hat a})_{\bar l}^i\,\bar\phi^{\bar l}\ , \\
   \cD_\mu\bar\phi^{\bj}&=\D_\mu\bar\phi^{\bj}
      -\frac{i}{2}\partial_\mu{\bar z}^{\hat a}(\bar\chi_{\hat a})_l^{\bj}\,\phi^l\  .
\end{split}
\end{equation}
Note that the definition of the covariant derivative $\cD$ contains a connection of the normal bundle of the D3-brane, a connection of the gauge group $U(N)$, and finally the newly added connection to include complex structure fluctuations. 
Inserting \eqref{gdefo} and \eqref{eq:Vectpertderiv} 
into the kinetic term of (\ref{eq:DBIlowone})
results in a cancellation of the terms proportional to
$z$ leaving only terms
proportional to $\partial_\mu z$ inside the covariant
derivatives.
We obtain up to linear order in $z$\footnote{In the covariant 
derivatives we dropped the term quadratic in $z$.} 
\begin{equation}\label{Lkinn}
   \mathcal{L}_{\text{kin}}
=-\frac{6\mu_3}{\Kw}\, \tr g_{i\bj}
     \cD_\mu\phi^i\cD^\mu\bar\phi^{\bj} 
=\frac{6i\mu_3}{\Kw}\, \tr v^\alpha(\omega_\alpha)_{i\bj}
     \cD_\mu\phi^i\cD^\mu\bar\phi^{\bj}\ ,
\end{equation}
where we also used \eqref{transJ}.  
Thus by going to a complex basis as is required
by supersymmetry, additional (derivative)
couplings between the complex structure deformations $z$ and
the matter fields $\phi$ arise.
As we will see
in section 3 these couplings also induce 
additional couplings in the K\"ahler potential
which destroy the product structure of the  moduli space
as indicated in \eqref{modulispace}.

Similarly, we rewrite the quartic non-Abelian term
according to
\begin{equation}\label{phi4}
   g_{op}g_{mn} \com{\phi^o}{\phi^m}\com{\phi^n}{\phi^{p}}
\rightarrow
    2  g_{i\bj}g_{k\bar l} 
             \com{\phi^i}{\phi^k}\com{\bar\phi^{\bar l}}{\bar\phi^{\bj}}
         + 2 g_{i\bj}g_{k\bar l} 
             \com{\phi^k}{\bar\phi^{\bj}}\com{\phi^i}{\bar\phi^{\bar l}} \  .
\end{equation}
Note that there is no complex structure dependence in these 
terms because it is of higher order (5th order) in the fields
$\phi$ and $z$.
Substituting (\ref{Lkinn}) and (\ref{phi4})
into the action (\ref{eq:DBIlowone}), we arrive at 
\begin{equation}  \label{eq:DBIlow}
\begin{split}
   S_{\mathrm{DBI}}^{\mathrm{E}}
     =-\mu_3 \int_\WV d^4\xi&\sqrt{-g_4} 
     \ \tr\left(\frac{36\:e^{4A}}{\Kw^2}(1+\ell^2\Riem{n}{\tau}{\tau}{m}\phi^n\phi^m)
       +\frac{\ell^2}{4}e^{-\phi} F^{\mu\nu}F_{\mu\nu} \right. \\
      -&\frac{6i\ell^2 }{\Kw}\, v^\alpha(\omega_\alpha)_{i\bj}
        \cD_\mu\phi^i \cD^\mu\bar\phi^{\bj} 
       +\frac{9\ell^2e^{4A}}{\Kw^2}\,e^{\phi}(G^{(3)}-\bar G^{(3)})_{lmn} 
        {\phi^l}{\phi^m}{\phi^n} \\
      +&\frac{18\:\ell^2}{\Kw^2}\, e^{\phi}\left.  
        \left(g_{i\bj}g_{k\bar l} 
             \com{\phi^i}{\phi^k}\com{\bar\phi^{\bar l}}{\bar\phi^{\bj}}
         +g_{i\bj}g_{k\bar l} 
             \com{\phi^k}{\bar\phi^{\bj}}\com{\phi^i}{\bar\phi^{\bar l}} \right)   
             \vphantom{\frac{e^{4A}}{\Kw^2}}\right)\ .
\end{split}
\end{equation}

Our next task is to expand the Chern-Simons action 
\eqref{eq:ChernSimon}. 
It contains interior multiplications with $\phi$ acting on 
forms which in local coordinates is given by 
\eqref{eq:intmul}.\footnote{This makes sense as the $\phi$ are really elements of the normal space, namely for D3-branes, of the tangent space of the Calabi-Yau at the brane location $y_0$.}    If the $\phi$ were commuting quantities, the interior multiplication of a form with $\ins{\phi}\ins{\phi}$ would always yield zero.\footnote{The interior 
multiplication is an anti-derivation of degree $-1$.}
This, however, is not the case 
as the $\phi$ are non-Abelian. From a physics point of view the reduction of the gauge group $U(N)$ to $U(1)$, i.e.\ 
the transition from non-Abelian $\phi$ to Abelian $\phi$, corresponds to reducing the non-Abelian Chern-Simons action \eqref{eq:ChernSimon} to the Abelian Chern-Simons action. Therefore all terms involving $\ins{\phi}\ins{\phi}$ in \eqref{eq:ChernSimon} must vanish for the gauge group $U(1)$, because there are no such terms in the Abelian Chern-Simons action.
There is, however, a remnant of the disappearing  
$\ins{\phi}\ins{\phi}$-terms in the non-Abelian case, because 
these non-Abelian terms appear in the trace and if $\phi$ commutes with the form to be multiplied with, then we can use the cyclic property of the trace to rotate the order of the $\phi$. The result differs for an odd and even number of $\ins{\phi}$ operating on forms. 
In the former case the cyclic property generates an even permutation of $\phi$ and the interior multiplication yields a possibly non-vanishing term, however, in the latter case the rotation of $\phi$ corresponds to an odd permutation and due to the skew-symmetry of forms the expression vanishes. In summary we find for all integers $k$
\begin{equation}
  \tr\left(\ins{\phi}^{2k}C^{(q)}\right)=0 \quad\text{for}\quad \com{\phi}{C^{(q)}}=0 \  .
\end{equation}


In the reduction we expand the four-form $C^{(4)}$ 
around its background value which is determined by eqs.\
(\ref{F5}) and \eqref{e4A=alpha}. However, we do not
necessarily  want to insert this background immediately 
but rather allow small deviations. This is related to the fact 
that we introduced a potential in the bulk action
which determines the allowed background values for the fluxes
and the background values for the fields
only after minimization. In this way we will be able to
also include IASD fluxes into our analysis.
One way to include small deviations from its background value 
into the reduction of the D-brane action
is to expand the function $\alpha$ in the four-form potential 
according to 
\begin{equation}\label{defh}
\alpha=e^{4A} - h(y)\ .
\end{equation}
We expand the pull-back of the function $h$ to quadratic
order in $\phi$
\begin{equation}
\Em^*(h)=h(y_0)+\ell\nabla_n h|_{y_0}\,\phi^n+
                \frac{1}{2}\ell^2\nabla_m\nabla_n h|_{y_0}\,
\phi^n\phi^m 
       =\frac{1}{2}\ell^2 h_{nm}\phi^n\phi^m \  ,
\end{equation}
where  we use
\begin{align}
   h(y_0)=0\ , && \nabla_n h|_{y_0}=0\ , && 
h_{nm}\equiv\nabla_m\nabla_n h|_{y_0}\  .
\end{align}
This expresses the fact that we are expanding around 
the solution $\alpha=e^{4A}$. Furthermore,
using a complex basis one has 
$\Em^*(h)= \frac{1}{2}\ell^2 (h_{i\bj}\phi^i\bar\phi^{\bj} +
 h_{ij}\phi^i\phi^j +{\rm h.c.})$.

The additional piece of information we need 
in order to reduce the Chern-Simons action 
is the  pull-back formula \eqref{eq:ppbcomp} and as before
also the non-Abelian Taylor expansion formula 
\eqref{eq:nonabelianTaylor}.
Furthermore,  the trace in \eqref{eq:ChernSimon} 
has to be taken as a symmetrized trace over appropriate 
non-commuting quantities. Altogether this yields the expanded Chern-Simons action 
in the Einstein frame
\bea \label{eq:WZaction}
   S_{\mathrm{CS}}^{\mathrm{E}}\!\!\!
   & = &\!\!\! \mu_3\int_\WV\!\! d ^4\xi\sqrt{-g_{4}} \,
       \tr\!\left(\frac{36\:e^{4A}}{\Kw^2}\, 
\big(1+\ell^2\Riem{n}{\tau}{\tau}{m}\phi^n\phi^m\big)
-\frac{18\ell^2}{\Kw^2}\,  \big(h_{i\bj}\phi^i\bar\phi^{\bj} +
 h_{ij}\phi^i\phi^j +{\rm h.c.}\big)       \right)\nn \\
     &&-\ \mu_3\int_\WV d ^4\xi\sqrt{-g_{4}}\, \tr\left(\frac{9i\ell^2e^{4A}}{\Kw^2}\,
           e^\phi (*_6G^{(3)}+*_6\bar G^{(3)})_{nmp}\phi^n\phi^m\phi^p \right) \\
&&+\ \frac{\mu_3\ell^2}{4}\int_\WV\tr\left(\phi^i\cD_\mu\bar\phi^{\bj}
 -\bar\phi^{\bj}\cD_\mu\phi^i\right)(\omega_\alpha)_{i\bj}
            d x^\mu\wedge d D_{(2)}^\alpha 
    \ +\ 2 l\,\tr\left(F\wedge F\right) \ ,\nn
\eea
where we have integrated by parts the pull-back of the fluctuations of the 
$4$-form field in the expansion \eqref{exp1}.  

Now we can combine
the Dirac-Born Infeld 
action \eqref{eq:DBIlow} and the Chern-Simons action 
\eqref{eq:WZaction} to arrive at
\begin{equation}
\begin{split}  \label{eq:actionD3}
   S_{bos}^{\mathrm{E}}
   =&-\frac{\mu_3\ell^2}{4} \int_\WV d^4\xi \sqrt{-g_4} \ 
     \tr e^{-\phi} F^{\mu\nu}F_{\mu\nu}
     +\frac{\mu_3\ell^2}{2}\int_\WV l\,\tr\left(F\wedge F\right) \\
&+\mu_3\ell^2 \int_\WV d^4\xi\sqrt{-g_4} \ 
     \tr\left(\frac{6i }{\Kw}\, 
            v^\alpha(\omega_\alpha)_{i\bj}
            \cD_\mu\phi^i \cD^\mu\bar\phi^{\bj} 
-\frac{18}{\Kw^2}\,  \big(h_{i\bj}\phi^i\bar\phi^{\bj} +
 h_{ij}\phi^i\phi^j +{\rm h.c.}\big) 
\right) \\
 &+\frac{\mu_3\ell^2}{4}\int_\WV\tr\left(\phi^i\cD_\mu\bar\phi^{\bj}
        -\bar\phi^{\bj}\cD_\mu\phi^i\right)(\omega_\alpha)_{i\bj}
             dx^\mu\wedge dD_{(2)}^\alpha \\
&-\mu_3\ell^2\int_\WV d ^4\xi\sqrt{-g_{4}}\, \tr\left(\frac{9ie^{4A}}{\Kw^2}\,
e^\phi (*_6G^{(3)}-i G^{(3)})_{nmp}\phi^n\phi^m\phi^p 
+ {\rm h.c.}\right. \\
&\qquad\qquad\qquad+\frac{18}{\Kw^2}\, e^{\phi}\left.  
        \left(g_{i\bj}g_{k\bar l} 
             \com{\phi^i}{\phi^k}\com{\bar\phi^{\bar l}}{\bar\phi^{\bj}}
         +g_{i\bj}g_{k\bar l} 
             \com{\phi^k}{\bar\phi^{\bj}}\com{\phi^i}{\bar\phi^{\bar l}} \right)   
             \vphantom{\frac{e^{4A}}{\Kw^2}}\right)   \ .
\end{split}   
\end{equation}
Note that the 
the first terms in \eqref{eq:DBIlow} and  \eqref{eq:WZaction}
exactly canceled each other. This is a consequence
of the fact that we are expanding around a
consistent background determined by
(\ref{e4A=alpha}). The deviation from this background is captured by 
the mass terms $h_{nm}\phi^n\phi^m$.
The term trilinear in the $\phi^n$
vanishes in the Abelian limit and also for
ISD-fluxes. 

This action has to be added to the bulk action 
\eqref{Bulk1} in appendix \ref{KKreduction}
and the self duality of the (modified) five-form field strength 
has to be imposed. This eliminates the two forms $D_{(2)}^\alpha$ in favor of
the scalars $\rho_\alpha$, leading to a modification of the  
derivatives of $\partial_\mu\rho_\alpha$ according to
\begin{equation}
   \partial_\mu\rho_\alpha\rightarrow
\partial_\mu\rho_\alpha
      +\ell^2(\omega_\alpha)_{i\bj}\,\tr\left(\bar\phi^{\bj}\cD_\mu\phi^i-
          \phi^i\cD_\mu\bar\phi^{\bj}\right) .
\end{equation}
This covariant derivative was also introduced in \cite{FP}, where it was argued to come from a modified five-form Bianchi identity due to the source term of the charged D3-branes. In our analysis, it appears naturally through the expansion of the Chern-Simons action, which describes the couplings of the RR-charges of the branes to the bulk fields. Furthermore, in our case the covariant derivative
also includes couplings to the complex structure deformations.

The final chore is to rewrite the trilinear terms in terms of
complex variables.
In order to do so, we use the decomposition \cite{DWG}\footnote{%
The combination $e^{4A}*_6 G_3 -i \alpha G_3$ is closed when 
$\tau$ is constant. Up to to second order in $\phi$ we can set 
$\alpha=e^{4A}$,  which makes $e^{4A}(*_6 G_3 -i G_3)$ 
closed and consequently coclosed. Therefore, 
it can be expanded in harmonic 
three-forms of the warped Calabi-Yau manifold.
On a six-dimensional manifold they coincide with the 
harmonic three-forms of the unwarped Calabi-Yau.} 
\begin{equation}
e^{4A}(*_6 G_3 -i G_3)= \frac{2i}{w} \left(\G\, \Omega 
+ \Gw^{\hat{a} {\hat b}}_w \,\, {\bar \chi}_{\hat{b}}
\CHI_{\hat{a}}\right) \ ,
\label{expG3}
\end{equation}
where $\G,\CHI_{\hat{a}}$ are defined in eq.\ \eqref{susyfromW} and 
${\Gw_{w}}_{\hat{a} {\hat b}}$ is the warped version of
the metric defined in \eqref{metrics}. i.e.\
\beq \label{defw}
{\mathcal{G}_{w}}_{\hat a \hat b} \equiv 
  - \frac{1}{w}\int e^{-4A}\chi_{\hat a} \wedge \bar \chi_{\hat b}\ ,
\qquad w\equiv\int e^{-4A} \Ox \wedge \Oxb \ .
\eeq
We should apply the substitution rule \eqref{eq:Vectpertcomp} and, in addition, 
we need to take into account the complex structure dependence of $\Omega$ and $\chi_{\ha}$ in \eqref{expG3}. According to \cite{Candelas:1990pi}\footnote{The expression differs from our conventions by an $i$ compared to ref.\ \cite{Candelas:1990pi}.}
\beq\label{derivomega}
   \frac{\partial\Omega}{\partial z^{\hat a}}=k_{\hat a}\Omega + i \chi_{\hat a}\ , \qquad
   \frac{\partial\chi_{\hat a}}{\partial z^{\hat b}}=k_{\hat b}\chi_{\hat a}+
   \kappa_{\hat a\hat b}^{\hat c}\bar\chi_{\hat c}\ ,
\eeq
where $\kappa_{\hat a\hat b}^{\hat c}$ is defined in \cite{Candelas:1990pi}
but here we do not need its precise form.
With this in mind, we get the following substitution formula for the $(3,0)$ piece of the flux
\begin{equation}
\begin{split}
   \frac{1}{3!}\Omega_{ijk}\phi^i\phi^j\phi^k 
     \rightarrow & \frac{1}{3!}(1+k_{\hat a}z^{\hat a})\Omega_{ijk}
        (\phi+z^{\hat a}\chi_{\hat a}\bar\phi)^i(\phi+z^{\hat a}\chi_{\hat a}\bar\phi)^j
        (\phi+z^{\hat a}\chi_{\hat a}\bar\phi)^k \\
     &+\frac{i}{2!}z^{\hat a}(\chi_{\hat a})_{ij\bar k}
        (\phi+z^{\hat a}\chi_{\hat a}\bar\phi)^i(\phi+z^{\hat a}\chi_{\hat a}\bar\phi)^j
        (\bar\phi+\bar z^{\hat a}\bar\chi_{\hat a}\phi)^{\bar k} \ ,
\end{split}
\end{equation} 
which simplifies to linear order in $z$ to
\begin{equation}
   \frac{1}{3!}\Omega_{ijk}\phi^i\phi^j\phi^k
      \rightarrow \frac{1}{3!}(1+k_{\hat a} z^{\hat a})\Omega_{ijk}\phi^i\phi^j\phi^k \ .
\end{equation}
Analogously we obtain linearly in $z$ for the $(2,1)$ piece of the flux
\begin{equation}
   \frac{1}{2!}(\chi_{\hat a})_{ij\bar k} \phi^i\phi^j\bar\phi^{\bar k}
     \rightarrow 
        \frac{1}{2!}(1+k_{\hat b}z^{\hat b})(\chi_{\hat a})_{ij\bar k}
        \phi^i\phi^j\bar\phi^{\bar k}
        +\frac{1}{2!}\eta_{\hat a\hat b}^{\hat c}z^{\hat b}
           (\bar\chi_{\hat c})_{\bar\imath\bj k}\bar\phi^{\bar\imath}
           \bar\phi^{\bj}\phi^k+\frac{1}{3!}\zeta_{\hat a\hat b}\bar z^{\hat b}
           \Omega_{ijk}\phi^i\phi^j\phi^k  ,
\end{equation}
where $\eta_{\hat a\hat b}^{\hat c}$ and $\zeta_{\hat a\hat b}$
are combinations of $\chi$ and $\kappa$ appropriately contracted
but again the precise form is not relevant in the following.

Thus we obtain for the trilinear coupling to first order in $z$
\bea\label{strange}
\cL_{\phi^3} &\sim& 
\G\, (1+k_{\hat a} z^{\hat a})
          \Omega_{ijk}\phi^i\phi^j\phi^k \\
    &+&\mathcal{G}_w^{\hat a\hat b}
\CHI_{\hat a}
      \left( 3(1+\bar k_{\hat c}\bar z^{\hat c})
          \bar\chi_{\hat b\:\bar\imath\bj k}
          \bar\phi^{\bar\imath}\bar\phi^{\bj}\phi^k
          +3\bar{\eta}^{\hat c}_{\hat a\hat b}\bar z^{\hat b}
          \chi_{\hat c\:ij\bar k}\phi^i\phi^j\bar\phi^k 
    +\bar \zeta_{\hat b\hat c}z^{\hat c}
             \bar\Omega_{\bar\imath\bj\bar k}
             \bar\phi^{\bar\imath}\bar\phi^{\bj}\bar\phi^{\bar k}\right)+\text{h.c.}\nonumber
\eea
where we used \eqref{expG3}. 
Finally, we insert
\eqref{strange} into  \eqref{eq:actionD3} to
obtain
\begin{equation}
\begin{split} \label{eq:fullD3action}
   S_{\mathrm{D}3}^{\mathrm{E}}
   =&-\frac{\mu_3\ell^2}{4} \int_\WV d^4\xi \sqrt{-g_4} \ 
     \tr e^{-\phi} F^{\mu\nu}F_{\mu\nu}
     +\frac{\mu_3\ell^2}{2}\int_\WV l\,\tr\left(F\wedge F\right) \\
&+\mu_3 \ell^2\int_\WV d^4\xi\sqrt{-g_4} \ 
     \tr\left(\frac{6i}{\Kw}\, 
            v^\alpha(\omega_\alpha)_{i\bj}
            \cD_\mu\phi^i \cD^\mu\bar\phi^{\bj} 
-\frac{18}{\Kw^2}\,  \big(h_{i\bj}\phi^i\bar\phi^{\bj} +
 h_{ij}\phi^i\phi^j +{\rm h.c.}\big) 
\right) \\
    &+\frac{\mu_3\ell^2}{4}\int_\WV\tr\left(\phi^i\cD_\mu\bar\phi^{\bj}
            -\bar\phi^{\bj}\cD_\mu\phi^i\right)(\omega_\alpha)_{i\bj}
            dx^\mu\wedge dD_{(2)}^\alpha \\
    &-\mu_3\ell^2\int_\WV d ^4\xi\sqrt{-g_{4}}\, 
    \tr\Big[ \frac{18}{\Kw^2}\, e^{\phi}  
        \left(g_{i\bj}g_{k\bar l} 
             \com{\phi^i}{\phi^k}\com{\bar\phi^{\bar l}}{\bar\phi^{\bj}}
         +g_{i\bj}g_{k\bar l} 
             \com{\phi^k}{\bar\phi^{\bj}}\com{\phi^i}{\bar\phi^{\bar l}} \right)  \\
    &\qquad -\frac{18e^{\phi}}{w\Kw^2} \Big(
     \G (1+k_{\hat a}z^{\hat a}) \Omega_{ijk}\phi^i\phi^j\phi^k 
          +3\mathcal{G}_w^{\hat a\hat b} 
     \CHI_{\hat a}
           \Big( (1+\bar k_{\hat c}\bar z^{\hat c})
           (\bar\chi_{\hat b})_{\bar\imath\bj k}
           \bar\phi^{\bar\imath}\bar\phi^{\bj}\phi^k\\
    &\qquad\qquad+ 3i
           \bar{\eta}^{\hat c}_{\hat a\hat b}\bar z^{\hat b}
           (\chi_{\hat c})_{ij\bar k}
           \phi^i\phi^j\bar\phi^k  
           + \bar \zeta_{\hat b\hat c}z^{\hat c}
           \bar\Omega_{\bar\imath\bj\bar k}
           \bar\phi^{\bar\imath}\bar\phi^{\bj}\bar\phi^{\bar k}\Big)+\text{h.c.}\Big)\Big]\ . 
\end{split}   
\end{equation}
(Note that $\mu_3\ell^2 = \tfrac1{2\pi}$ holds for D3-branes.)

This concludes our derivation of the bosonic action
of a stack of $N$ space-time filling D3-branes 
in a Calabi-Yau orientifold compactification.
We now turn to its fermionic counterpart.

\subsubsection{The fermionic action}\label{subsubsec:D3fermions}

The supersymmetrized version of the Abelian DBI-CS action has been proposed by different groups 
\cite{superaction}. 
The supersymmetric action looks exactly like the bosonic one, but in the former case the brane lives in superspace, 
and every space-time field is promoted to a superfield. The explicit form of the action is given in terms of a
background supervielbein and super NS-NS and R-R gauge fields, that have to be expanded in terms of
component fields. The action has a fermionic kappa-symmetry, which
allows to project out half of the fermions by a gauge choice
 and is obviously reparametrization invariant. 
When choosing static gauge to fix this latter symmetry, the space-time fermions become
world-volume fermions, and these are the fermionic world-volume degrees of freedom. 

The non-Abelian version of the supersymmetric action is not known yet. We expect it 
to contain a trace over the gauge indices of the corresponding Abelian expression plus
eventually extra terms. Any extra term must have dimension greater than three,
so up to dimension three, which is all we need to compute fermionic supersymmetric and gaugino soft masses, there should
not be any additional intrinsically non-Abelian terms.

The terms in the Abelian D3-brane action containing fermion bilinears were found in \cite{D3}. We  write this action 
in a fixed gauge for the kappa-symmetry, 
adding a trace over the gauge indices. In the four-dimensional
Einstein frame the Lagrangian is\footnote{Our conventions differ from 
those in \cite{MG} by a factor of $i$ in the kinetic term.}  
\beq
\cL^{\rm E}_{\rm ferm}= \frac{36 \mu_3}{\Vw^2} \
\tr \left( -\frac{i\Vw^{1/2}e^{3A}}{2 \sqrt 6} \
\overline{\theta} {\Gamma}^{\mu} \D_{\mu} \theta 
+  \frac{e^{{\phi}/2}e^{4A}}{48}\  {\rm Re}
[ (*_6 G^{(3)}- i G^{(3)})_{pqr}]\overline{\theta }\Gamma^{pqr}\theta \right)
\label{fermlagrangian}
\eeq
where 
the covariant derivative $\D_{\mu}$ contains a spin connection and, in the non-Abelian generalization
of the fermionic action, it should also contain the connection of the
gauge group $U(N)$. 

The spinor $\theta$ is a $D=10$ Majorana-Weyl fermion of negative chirality, i.e.\ it is in the ${\bf 16}$ of
$SO(9,1)$. Under $SO(9,1)\rightarrow SO(3,1) \times SO(6)$, it decomposes into $({\bar {\bf 2}}, {\bf 4}) 
\oplus ( {\bf 2}, {\bar {\bf 4}})$. This implies that from the 
four-dimensional point of view, there are
four fermions: the gaugino $\lambda$, which together with $A_{\mu}$ forms an ${\cal N}=1$ vector multiplet, 
and three fermions $\psi^i$, which are the superpartners of the scalars $\phi^i$, resulting in three ${\mathcal N}=1$
chiral multiplets.  
Ten-dimensional gamma matrices also decompose as
\beq
{\Gamma}^{\mu}={\gamma}^{\mu} \otimes 1 \, , \qquad \Gamma^m=\gamma_5 \otimes \gamma^m
\eeq
where ${\gamma}^{\mu}$ and $\gamma^m$ are Dirac gamma matrices of $SO(3,1)$ (in four-dimensional Einstein frame) and $SO(6)$ respectively.

We want to write the action in terms of the four-dimensional gaugino $\lambda$ and the fermions $\psi^i$, 
so we need to see how these fermions are ``hidden'' in $\theta$. This decomposition is carried out in detail in 
Appendix \ref{ap:spinor}. Here we quote the result:
\beq
\theta= \ell \,e^{-3A/2}\,\left( \left(\frac{\Vw}{6}\right)^{1/4}\frac{1}{2 \|\Omega\|}\,\Omega_{ijk}\, \psi^i \otimes
{\gamma}^{jk}\, \chi 
+ \left(\frac{\Vw}{6}\right)^{3/4}\, e^{-\hphi/2}\,  \lambda \otimes \chi \right) + {\rm h.c.}\ .
\label{thetadecomp}
\eeq
Inserting this decomposition in the action
we get the following kinetic terms\footnote{We also get terms of the form 
$\overline{\lambda} {\gamma}^{{\mu}} 
 \lambda \,\partial_{\mu} \Vw$, for example, but these are the same order as terms we have already neglected in the action, 
so we also neglect them here.}
\beq
{\cal L}_{\rm kin}= -i \mu_3 \ell^2 \tr \left( \frac{6}{\Vw}\
\overline{\psi}^{\ib}{\gamma}^{\mu} 
\D_{\mu} \psi^l {g}_{\ib l} + e^{-\hphi} \,\overline{\lambda}
{\gamma}^{{\mu}} 
\D_{\mu} \lambda \right)\ .\label{Lkin}
\eeq
This is the correct supersymmetrization of the kinetic terms in the bosonic action (\ref{eq:fullD3action}).

The interaction term in (\ref{fermlagrangian})
contains the combination $e^{4A}(*_6 G^{(3)} -i G^{(3)})$ that
appeared already in the bosonic action, and contains only the IASD pieces of 3-form flux. 
Inserting (\ref{expG3}) in the second term of (\ref{fermlagrangian}), we get the following interaction Lagrangian\footnote{We should also insert 
here the expansion of $\Ox$ and $\chi_{\ha}$ to first order in the
complex structure deformations, as done in the bosonic action, but this will give additional terms of the same order
of those we have neglected in the fermionic action.} 
\beq
{\cal L}_{\rm int}= \frac{3i\mu_3 e^{\hphi/2}}{4\,\Vw^2\,w\,}   
\left[ 
\G\, \Omega_{ijk}  {\overline \theta} \Gamma^{ijk} \theta\ 
+ 3\, \Gw^{\hat{a} {\hat{ b}}}\, 
\CHI_{\hat{a}} ({\bar \chi}_{\hat{b}})_{i\bj\bk}\,{\overline \theta}
  \Gamma^{i\bj \bk} \theta \right] + {\rm h.c.}
\ .
\eeq
The first contraction in this equation is
\beq
\Omega_{ijk}{\overline \theta} \Gamma^{ijk} \theta\, = 
\frac{1}{6 \sqrt6} \, \ell^2 \, \Vw^{3/2} e^{-\phi} \lambda \lambda \Omega_{ijk}\,\chi^T {\gamma}^{ijk} \chi\, 
= \frac{1}{\sqrt6}\,\ell^2\, \, \|\Omega\| \Vw^{3/2} e^{-\phi} \lambda \lambda
\eeq
where in the last equality we have used (\ref{omega}). 
Similarly, the second contraction gives
\beq
 ({\bar \chi}_{\hbb})_{ i\bj\bk}{\overline \theta} \Gamma^{i\bj
   \bk} \theta\, =- 3 \sqrt 6 \ell^2 \Vw^{1/2} 
\frac{1}{\|\Omega\|} 
\,({\bar \chi}_{\hbb})_{\,i\bk \bl}\, \Omega^{{\bar s}\bk \bl}
{g}_{j {\bar s}}\, \psi^i \psi^j = - 3 \sqrt 6 \ell^2 \Vw^{1/2} \|\Ox\| (\chi_{\hbb})_i^{\bar s} {g}_{j {\bar s}}\, \psi^i \psi^j 
\eeq
So we get for the interaction Lagrangian
\beq
{\cal L}_{\rm int}= -\frac{1}{4} \mu_3 \ell^2 e^{\hat K/2}  
\left[ \frac{e^{-\phi}}{2}
\G\, \lambda \lambda \,  
 -   \frac{27}{\Vw} \,\CHI_{\hat{a}}\, \Gw^{\hat{a} \hbb} \,({\bar
  \chi}_{\hbb})_{i}^{\bar s}\,{g}_{j {\bar s}} \,
 \psi^i \psi^j\right]  + \,\,  {\rm h.c.}\ , 
\label{Lint}
\eeq
where we use
\beq
\|\Omega\|=\sqrt{\frac{6iw}{\Kw}}\ ,\qquad
e^{\hat K/2}= \frac{\sqrt{18i} e^{\hat \phi/2}}{\Vw \sqrt w}\ .
\label{relwo}
\eeq
This completes our computation of the fermionic kinetic 
and mass terms. Their consistency with the bosonic Lagrangian 
we are going to check in the next section.


\section{Soft Supersymmetry Breaking Terms}\label{sec:ssb}


So far we computed the low energy effective action of type IIB 
Calabi-Yau orientifold compactification with background D-branes 
and fluxes. Our next 
task is to rewrite this action
in a standard supergravity form. Since 
supersymmetry is spontaneously broken by the three-form fluxes
it is convenient
to use instead of the action \eqref{N=1action_WB},
one that is already adopted to this situation.
The case of $F$-term supersymmetry breaking
by moduli fields has been analyzed generically in
refs.\ \cite{KL,BIM}. 
In the limit $M_{\rm Pl}\to \infty$ with 
the gravitino mass $m_{3/2}$ fixed, the resulting
effective action corresponds to a softly broken, globally
supersymmetric theory which is characterized
by a K\"ahler potential $K$, a superpotential $W^{\eff}$,
a gauge kinetic function $f$
and a set of soft supersymmetry breaking terms.
After briefly recalling the notation and results of \cite{KL,BIM}
in section \ref{subsec:sugra}, we determine in sections 
\ref{subsec:softD3}--\ref{subsec:IASD}
the supersymmetric couplings and the soft supersymmetry breaking
terms by comparing the generic action of
\cite{KL,BIM} with the action computed for
D3-branes in a Calabi-Yau orientifold bulk in the previous section.
The internal consistency of the derived formulas is 
highly non-trivial check on the results of the previous section.

\subsection{Soft Supersymmetry Breaking from Supergravity}\label{subsec:sugra}

In ${\mathcal N}=1$ supergravities arising from compactification of string
theory it is convenient to distinguish between the chiral charged
matter fields $\phi^i$ and the gauge neutral moduli $\M^I$.
(Here $\phi^i$ and $\M^I$ denote the bosonic (lowest)
components of chiral multiplets.)
In terms of the fields introduced in the previous section
this corresponds to the identification 
$\M^I = (T_\alpha,z^{\hat a},G^a,\tau)$
while $\phi^i$ are the same charged fields as introduced in 
section \ref{subsec:D3action}.
As long as the gauge symmetry is unbroken, the vacuum expectation value (VEV)
of the $\phi^i$ vanishes and  therefore it is  convenient\footnote{%
Here we follow the analysis of refs.\ \cite{KL,BIM} 
where the notation is adjusted to our situation.
We also choose to set $M_{\rm Pl} = 1$.}
 to expand 
the K{\"a}hler potential $K(\M^I,\bar\M^I, \phi^i,\bar\phi^i)$ 
and the superpotential $W(\M^I,\phi^i)$ in a power series in
$\phi^i$ 
\bea\label{KWexp}
K(\M,\bar\M, \phi,\bar\phi) &=& \hat K(\M,\bar\M)
+ Z_{i\bar j}(\M,\bar\M)\, \phi^i\bar\phi^{\bar j}
+ \big(\frac12\, H_{ij}(\M,\bar\M)\, \phi^i\phi^{j} 
+ \textrm{h.c.} \big)
+ \ldots\ , \nonumber\\
W(\M,\phi) &=&\hat W(\M) 
+ \frac 12\, \tilde \mu_{ij}(\M)\, \phi^i\phi^{j}
+ \frac 13\, \tilde Y_{ijk} (\M)\, \phi^i\phi^{j}\phi^{k} +\ldots\ .
\eea
 The gauge couplings $g$
are only computed at tree level and therefore obey
\beq\label{gf}
g^{-2} = {\textrm Re} f(\M)\ ,
\eeq
where $f$ is the holomorphic gauge kinetic function.

Spontaneous supersymmetry breaking occurs if 
a $D$-term  or a $F$-term has a non-vanishing VEV.
As we already noted  in section \ref{subsubsec:fluxes},  
the three-form fluxes $G^{(3)}$ 
only lead to non-vanishing $F$-terms induced by 
the moduli dependent superpotential $\hat W(\M)$ 
given in \eqref{hatW}. 
Since the $\phi^i$ vanish in the ground state they do not
contribute to the  $F$-terms and one has
\beq\label{Fterm}
\bar F^{\bar I}\ =\  e^{\hat K/2}\,\hat K^{\bar I J} 
\left( \partial_J\hat W + \hat W \partial_J \hat K\right) \ ,\qquad
\eeq
where $\hat K^{\bar I J} \,=\,(\hat K_{\bar I J})^{-1}$. 
For vanishing cosmological constant the gravitino mass
\beq\label{m32}
m_{3/2}\ =\ e^{\hat{K}/2} \hat{W}
\eeq
 is an alternative measure 
for the supersymmetry breaking.

Without going through the analysis let us just state the resulting 
effective low energy supergravity potential in the standard
limit where one sends $M_{\rm Pl}\to \infty$
with $m_{3/2}$ fixed. In this limit one finds \cite{KL,BIM}
\bea\label{Veff}
V^{\eff} &=& \frac12 D^2
+  Z^{i\bj}\, (\partial_i \Weff)(\partial_{\jb} \Weffb)
+\ m^2_{i\bj,{\rm soft}}\, \phi^i\bar\phi^{\jb} \nonumber\\
&& 
+\frac13\, A_{ijk}\phi^i\phi^{j} \phi^k + 
\frac12\, B_{ij}\phi^i\phi^{j} + \textrm{h.c.}\ ,
\eea
where the `supersymmetric' terms are given by
\bea\label{Weff}
D &=& - g\ \bar\phi^\ib Z_{\ib j}\phi^j\ ,\nonumber\\
\Weff &=&\frac{1}{2}\, \mu_{ij}\, \phi^i \phi^j 
+\frac{1}{3}\, Y_{ijk}\, \phi^i \phi^j \phi^k \ ,\\
\mu_{ij}&=& e^{\hat{K}/2} \tilde{\mu}_{ij} + m_{3/2} H_{ij} - F^{\bar I} \bar{\partial}_{\bar I} H_{ij}\ ,\nonumber \\
Y_{ijk}&=& e^{\hat{K}/2} \tilde{Y}_{ijk}\ .\nonumber
\eea
$\Weff$ also determines the fermionic masses via
\beq
m^f_{ij}\ =\ \partial_i\partial_j \Weff\ =\mu_{ij}\ .
\eeq
The soft supersymmetry breaking terms read
\bea\label{softi}
m^2_{i \bj,{\rm soft}}&=& (|m_{3/2}|^2 +V_0)\, Z_{i \bj} - 
F^I F^{\bar J} R_{I {\bar J} i \bj}  ,\nonumber \\
A_{ijk} &=& F^I D_I Y_{ijk}\ , \\
B_{ij} &=& \big(2|m_{3/2}|^2 +V_0\big)\, H_{ij}
- \bar m_{3/2} \bar F^{\bar J}\bar\partial_{\bar J} H_{ij}
+  m_{3/2} F^{I} D_I H_{ij} \nn \\
&& - F^{I}\bar F^{\bar J}D_I\bar\partial_{\bar J} H_{ij}
- e^{K/2} \tilde \mu_{ij}  \bar m_{3/2}
+e^{K/2}F^{I}D_I\tilde \mu_{ij} \ ,\nonumber
\eea
where
\bea\label{curvi}
R_{I \bar J i \bj} &=& \partial_I \partial_J Z_{i \bj} - 
   \Gamma^{k}_{Ii}Z_{k \bar l}\Gamma^{\bar l}_{\bar J \bj} \ ,\qquad
\Gamma ^l_{Ii}=Z ^{l\bj} \partial_I Z_{\bj i}\ ,\nonumber\\
D_I Y_{ijk} &=& \partial_I Y_{ijk} + \frac{1}{2} \hat{K}_I Y_{ijk} -
3\Gamma ^l_{Ii} Y_{ljk} \,  ,\\
D_I \tilde \mu_{ij} &=& \partial_I \tilde\mu_{ij} + \frac{1}{2} \hat{K}_I \tilde\mu_{ij} -
2\Gamma ^l_{Ii} \tilde\mu_{lj} \ .   \nonumber
\eea
This form of the soft terms differs from the ones given in \cite{KL}
in three respects.
First of all, it includes
the possibility of a non-vanishing
cosmological constant $V_0=\langle V\rangle$. 
Secondly,
the $B$-term given in \cite{KL}
reads 
$B_{ij} = F^I D_I \mu_{ij} -  \bar m_{3/2}\mu_{ij}$
(for vanishing cosmological constant).
This equation can be obtained from the $B$-term given in 
\eqref{softi} by using the condition for the ground state
$\partial_I V = 0$ and a vanishing cosmological
constant $V_0=0$. However, for the purpose of this paper,
the form  of \eqref{softi} is more convenient in that 
we do want to allow for a cosmological constant but more
importantly later on
we compare the terms computed from the D3-brane
action with the supergravity formulas just given.
In this comparison  it is inconvenient
to impose $\partial_I V = 0$.
Thirdly,
in \cite{KL} it was assumed that Yukawa couplings
$Y_{ijk}$ and the supersymmetric mass terms $\mu_{ij}$ 
are symmetric
in its indices which led to slightly different formulas. 
Here we allow for arbitrary symmetries and changed the covariant
derivatives accordingly. 

Finally, for the gaugino masses one finds 
\begin{equation}\label{mgaugino}
  m_g = F^I \partial_I\,  \text{ln}( \text{Re} f) \ .
\end{equation}

The next step is to read off the couplings and masses
of this softly broken supersymmetric theory from the actions
\eqref{N=1action_WB}, \eqref{eq:fullD3action}, \eqref{Lkin}
and  \eqref{Lint}.

\subsection{The supersymmetric terms}\label{subsec:softD3}

In this section we extract the supersymmetric couplings, 
that is 
$Z_{ij}, H_{ij}, W^{\eff}$ and $f$, 
 by
comparing with the action computed in 
\eqref{N=1action_WB}, \eqref{eq:fullD3action}, \eqref{Lkin}
and  \eqref{Lint}. This analysis is then completed in the following
two sections where  the soft 
supersymmetry breaking terms
$m_{i \bj,\, {\rm soft}}, B_{ij}, A_{ijk}, m_g$
are determined.

Let us start with the gauge kinetic function $f(M)$
which can be straightforwardly determined from 
the first two terms of 
\eqref{eq:fullD3action} to be 
\beq \label{ftau}
f=-i \mu_3\ell^2\tau = -\frac{i}{2\pi}\, \tau\ ,
\eeq 
where we use $\mu_3\ell^2 = \tfrac1{2\pi}$.

Let us continue 
 with the kinetic terms and determine
the K\"ahler potential or rather $Z_{i\bj}$ and 
$H_{ij}$.
$\hat K(\M,\bar\M)$ was already determined in \eqref{kaehlerpot-O7-1}
and we should read off $Z$ and $H$ from \eqref{eq:fullD3action}.
Surprisingly  it is possible and in fact easier to compute the full
$K(\M,\bar\M, \phi,\bar\phi)$ and then do the expansion
as in \eqref{KWexp} rather than just reading off 
$Z_{i\bj}$ and 
$H_{ij}$ from \eqref{eq:fullD3action}.\footnote{In fact
it is straightforward to read off $Z_{i\bj}\sim g_{i\bj}$
from \eqref{eq:fullD3action} but $H_{ij}$ is more involved.}
The reason is 
the somewhat complicated field redefinition \eqref{tau}
which was
necessary in order to transform the action obtained from the 
Kaluza-Klein reduction to the standard $\cN=1$ form 
\eqref{N=1action_WB}. The same is true if one includes 
the kinetic terms of the matter fields $\phi^i$
computed in \eqref{eq:fullD3action}. One finds that 
the K\"ahler potential  takes the form 
\begin{equation} \label{eq:KPbrane}
 K(\tau,T,G,z,\phi)=-\ln\left[-i\int\Omega \wedge\bar\Omega \right] 
  -\ln\left[-i(\tau - \bar \tau)\right]
-2 \ln\Big[\tfrac16\cK(\tau,T,G,z,\phi)\Big] \ ,
\end{equation}
which strongly resembles the $\hat K$ given in 
\eqref{kaehlerpot-O7-1}. As before $\cK$ is proportional to 
the volume of the Calabi-Yau manifold, i.e.
$\cK \equiv \cK_{\alpha \beta \gamma} v^\alpha v^\beta v^\gamma$
still holds,
but the dependence of the $v^\alpha$ on the K\"ahler coordinates
 is modified in that they are now also a function of the 
complex structure moduli $z^{\hat a}$ and the brane fluctuations $\phi$. 
More precisely, the K\"ahler coordinates $(\tau,T,G,z,\phi)$
are defined  in close analogy with \eqref{tau}, and read\footnote{%
In this coordinate transformation we are only working to leading
(linear) order in $z^{\hat a}$.}
\bea\label{eq:KKc}
\tau&=&l+ie^{-\phi} \ ,\qquad\qquad
   G^a\ =\ c^a-\tau b^a  \ , \\
   T_\alpha&=&\frac{3i}{2}\rho_\alpha+\frac{3}{4}
             \cK_\alpha- \frac{3i}{4(\tau-\bar\tau)}\kappa_{\alpha bc}G^b(G-\bar G)^c 
+\frac{3}{2}i\mu_3\ell^2(\omega_\alpha)_{i\bj}\,\tr \phi^i \big(\bar\phi^{\bj}-\frac{i}{2}
   \bar z^{\hat a}(\bar\chi_{\hat a})^{\bj}_{ l}\phi^l\big) 
\ . \nn
\eea
Note that exactly as in \eqref{tau},
the functional dependence of $\cK$ 
in terms of the K\"ahler coordinates can only be stated implicitly
by solving \eqref{eq:KKc} for $v^\alpha$
in terms of $(\tau,T,G,z,\phi)$ and inserting into $\cK$.
Note also that now all moduli fields mix non-trivially 
in the K\"ahler potential \eqref{eq:KKc}, and  thus the 
K\"ahler metric is no longer block diagonal. In addition, $\cK$ 
 is also not of the sequestered form
suggested in \cite{RS} since $\cK$ generically
does not split into a sum
$\cK\neq \cK_{\rm hid}(\tau,T,G,z) + \cK_{\rm obs}(\phi)$.

For a single (radial) K\"ahler modulus $v$, \eqref{eq:KKc} can be solved
explicitly and one obtains 
\bea
  -2 \ln\; \cK &=& -3 \ln \frac{2}{3}\left[ T + \bar T
  + \frac{3i}{4(\tau - \bar \tau)} \cK_{1 a b} (G-\bar G)^a (G-\bar G)^b \right.\\
 &&  \left.\qquad + 3i \mu_3 \ell^2 (\omega_1)_{i \bj}\, \text{Tr} (\phi^i \bar \phi^{\bj})
  + \frac{3}{4}\, \mu_3 \ell^2 ((\omega_1)_{i \bj} \bar z^{\hat a} 
(\bar\chi_{\hat a})^{\bj}_{ l}\, \text{Tr} (\phi^i \phi^l)+h.c.)
  \right]\ . \nn
\eea
Note that in this case the K\"ahler potential is of the sequestered form 
if we also freeze the complex structure moduli $z^{\hat a} = 0$.
For $G^a = z^{\hat a} = 0$, a similar form of the K\"ahler potential was suggested
in \cite{DWG,KKLMMT}.  Here we derived it from a reduction
of the DBI-action and also included the couplings to
$G^a$ and the complex structure deformations  $z^{\hat a}$.

The full $K(\M,\bar\M, \phi,\bar\phi)$
can be expanded for small $\phi$ according to \eqref{KWexp}
which yields 
\beq \label{Z}
Z_{i \bj}= \frac{6 \mu_3 \ell^2}{\cK}\ {g}_{i\bj} 
= -i\frac{6 \mu_3 \ell^2}{\cK}\  v^{\alpha}(w_{\alpha})_{i \bj}\ ,
\eeq
and 
\beq
H_{ij}=\frac{3 \mu_3 \ell^2}{\cK \|\Omega\|^2} \, v^{\alpha} (w_{\alpha})_{(j| \bar s}\,{\Omega}^{{\bar s} \bk \bl}
(\bar \chi_{\hat{a}})_{|i)\bk \bl}\, {\bar
  z}^{\hat{{a}}}=\frac{3 \mu_3 \ell^2}{\cK} \, v^{\alpha} (w_{\alpha})_{(j| {\bar s}}\, ({\bar \chi}_{\hab})^{\bar s}_{|i)}
{\bar
  z}^{\hat{{a}}}\ .
\label{H}
\eeq

Next let us determine $W^{\eff}$. The Yukawa couplings
$Y_{ijk}$ are best determined from the quartic couplings 
in the potential while the $\mu$-terms
are best determined from the fermionic masses.
We start with quartic couplings which we read off from 
\eqref{eq:fullD3action}
\beq\label{Vquartic}
{V}^{\rm D3}_{\rm quartic}= 
    \frac{18\:\ell^2\mu_3e^{\phi} }{\cK^2}\ 
        \tr \left(g_{i\bj}g_{k\bar l} 
             \com{\phi^i}{\phi^k}\com{\bar\phi^{\bar l}}{\bar\phi^{\bj}}
         +g_{i\bj}g_{k\bar l} 
             \com{\phi^k}{\bar\phi^{\bj}}\com{\phi^i}{\bar\phi^{\bar l}} \right) \ .
\eeq
Using the Jacobi identity the two terms can be identified 
with the sum of a $D$-term and a term  derived from a superpotential.
Using eqs.\ (\ref{Veff}) and (\ref{Z}) 
we find\footnote{Since we have restricted our analysis to fourth
in the fields we do not see the holomorphic $z^{\hat a}$ dependence
in the quartic term \eqref{Vquartic}. Including such terms would result 
in $\Omega\to \Omega (1+k_{\hat a}z^{\hat a})$ 
where at linear order in $z^{\hat a}$ the function $k_{\hat a}$
is constant.}
\beq
Y_{ijk}= 3\mu_3 \ell^2  e^{\hK/2} \Ox_{ijk}\ ,
 \qquad  
{\tilde Y}_{ijk}= 3\mu_3 \ell^2 \Omega_{ijk} 
= \frac{3}{2\pi}\,  \Omega_{ijk} \ .
\label{tildeY}
\eeq

The supersymmetric $\mu$-terms
can be determined either from the fermionic mass terms of (\ref{Lint})
for the $\psi^i$ (which are the superpartners of the $\phi^i$) 
or from eq.\ \eqref{Weff}. As we are going to show 
both determinations agree up to numerical factors and
we find\footnote{Here we have taken the numerical factor from
the bosonic computation since it is under better control.}
\beq
\mu_{ij}^{\Dth}= -
\frac{\mu_3 \ell^2}{2w}\  e^{\hK/2}
\cG^{\hat{a} {\hbb}} \,\Omega^{{\bar s}\bk \bl}\, ({\bar
  \chi}_{\hbb})_{\,(i|\bk\bl}\,{g}_{|j) {\bar s}}
 \, \CHI_{\hat{a}} \ .
\label{muD3}
\eeq
We see that they are  proportional to 
IASD-fluxes
(through $\CHI_{\hat{a}}$) 
and thus vanish in the supersymmetric limit
(or more precisely for ISD-fluxes). 
$\mu_{ij}^{\Dth}$ has  to be  compared with 
the $\mu_{ij}$ of \eqref{Weff} which has two distinct pieces.
The first $\tilde \mu_{ij}$ term survives
in the supersymmetric limit while the second and third term are
proportional to $m_{3/2}$ and $F$, respectively,  which 
vanish in the supersymmetric limit. Thus, the
$\mu_{ij}^{\Dth}$ given in \eqref{muD3}
has to correspond to these latter terms or in other words is 
induced by a Giudice-Masiero
mechanism \cite{GM} and we conclude 
\beq
\tilde\mu_{ij}=0\ .
\eeq
This is consistent with the fact
that for vanishing fluxes all matter fields 
should be exactly massless. 
In order to check the consistency of \eqref{muD3}
with the supergravity expression \eqref{Weff}, 
we need to compute
\beq \label{muwithoutmu}
\mu_{ij}^{\eff}= m_{3/2} H_{ij} - F^{\bar I} \bar{\partial}_{\bar I} H_{ij}\ ,
\eeq
and show that it coincides with $\mu_{ij}^{\Dth}$ of
\eqref{muD3}.
{}From eq.\ \eqref{H} we learn that $
H$ depends explicitly on the complex structure moduli $z^{\hat a}$, 
and through $v^{\alpha}$ and $\Vw$ also 
implicitly on the other moduli.
Using the formulas given in appendix \ref{ap:F-termformulae}, we find
\beq \label{cancellationinmu}
 \bar F^{\bar I} \partial_{\bar I} H_{ij} 
= m_{3/2} H_{ij}- \bar F^{\bar z^{\hat a}} \partial_{\bar z^{\hat
    a}} H_{ij}\ .
\eeq
Inserted into \eqref{muwithoutmu}  yields
\beq
\mu_{ij}^{\eff} =  - \bar F^{\bar z^{\hat a}} \partial_{\bar z^{\hat
    a}} H_{ij}\ .
\label{mushort}
\eeq
{}From \eqref{H} we obtain  
\beq
\partial_{\bar z^{\hat{b}}} H_{ij}
= \frac{3 \mu_3 \ell^2}{\cK} \, 
v^{\alpha} (w_{\alpha})_{(j| {\bar s}}\,(\bar{\chi}_{\hat{{b}}})^{\bar s}\,_{|i)}
=\frac{ \mu_3 \ell^2}{2w}\, {g}_{(j| \bar{s}}\,{\Omega}^{{\bar s} \bk \bl}
(\bar{\chi}_{\hat{{b}}})_{|i)\bk \bl}\ ,
\eeq
which, when inserted into  \eqref{mushort} shows (up to numerical factors)
$\mu_{ij}^{\eff} = \mu_{ij}^{\Dth}$.

Let us close this subsection with the observation that
due to the presence of both $\mu_{ij}$ and $Y_{ijk}$
a cubic mixed term arises from 
the $Z^{i\bj}\, (\partial_i \Weff)(\partial_{\jb} \Weffb)$
of \eqref{Veff}.
Inserting $\mu_{ij}$ of 
eq.\ \eqref{muD3}  and the Yukawa
couplings $Y_{ijk}$ of (\ref{tildeY}), we get  
\bea\label{Vcubic}
V^{\eff}_{\rm cubic,\,susy}&=&  \frac{\cK}{9 \mu_3 \ell^2}\, \mu_{li}\, {\bar Y}_{\bj \bk 
  {\bar l}}\, {g}^{l \bj} \,\phi
  ^i \phi ^{\bk} \phi ^{\bar l} + {\rm h.c. }  \nn\\
&=&- \frac{ \mu_3 \ell^2}{6} e^{\hK} 
  \cG^{ \hat a \hat{b}} 
\CHI_{\hat a}
 \frac{1}{w} { g}_{(l| {\bar s}}\, {\bar
   \Omega}^{{\bar s} {\bar t} {\bar u}} \,({\bar \chi}_{\hat{b}})_{|i) {\bar
     t}{\bar u}} \,{\bar \Omega}_{\bj \bk {\bar l}} \, {g}^{l \bj}\,\, \phi
  ^i \phi ^{\bk} \phi ^{\bar l} + {\rm h.c} \\
&=& -3i \mu_3 \ell^2 \,e^{\hK}  \cG^{\hat{a} \hat{b}} 
\CHI_{\hat a}
(\bar{\chi}_{\hat{b}})_{i \bar{k} {\bar l}} \, \phi
  ^i \phi ^{\bk} \phi ^{\bar l} + {\rm h.c.}\nonumber
\label{trilineareffsusy}
\eea
which is indeed equal to the trilinear mixed term 
in \eqref{eq:fullD3action}.\footnote{One might have worried
that these mixed cubic terms are hard breaking terms since
they are proportional to the fluxes yet at the same time
they cannot be holomorphic $A$-terms. This `puzzle' is resolved by
the fact that they arise as mixed supersymmetric terms
of a Giudice-Masiero $\mu$-term with Yukawa-terms.}
This also is an independent check on the $\mu$-term given in 
\eqref{muD3}.

This completes the determination of the supersymmetric terms and 
we now turn to the soft terms.

\subsection{Soft supersymmetry breaking terms from supergravity
}\label{subsec:SSB}

In this section we compute the soft supersymmetric terms
from eqs.\ \eqref{softi}.
Let us start with the  soft mass terms. 
Using \eqref{usefull} we  derive
\beq \label{cancellationform}
F^I F^{\bar J} R_{I \bar J i \bj}= |m_{3/2}|^2 Z_{i \bj}\ .
\eeq
Inserted into (\ref{softi}) we arrive at 
\beq
m^{2 \,\,\eff}_{i \bj,{\rm soft}}= V_0 Z_{i \bj} \ ,
\label{msofteq}
\eeq
which corresponds to universal soft scalar masses.
For the $B$-terms 
we first compute using \eqref{usefull} and \eqref{Fgamma}
\beq \label{cancellationforB}
\left({\bar m}_{3/2} \bar F^{\bar I} \partial_{\bar I} +  m_{3/2} F^{I} D_I - F^I \bar F^{\bar J} D_I \partial_{\bar J} \right)H_{ij}=
-2 |m_{3/2}|^2 H_{ij}\ .
\eeq
Inserted into (\ref{softi}) and using $\tilde\mu=0$ we 
obtain
\beq \label{Beq}
B_{ij}= V_0\, H_{ij}\ .
\eeq
Note that this $B$-term is not proportional to the 
$\mu$-term \eqref{muD3}.
For the gaugino masses we insert \eqref{ftau} 
and \eqref{F^A} into \eqref{mgaugino} to arrive at\footnote{%
The  masses for the bulk gauginos can be computed 
using \eqref{last}.}
\beq
m^{\eff}_g= -\frac{i}{2}\, e^{\phi} F^{\tau} = - e^{\hat K/2}\, \G\ .
\label{mshort}
\eeq
For the trilinear $A$-terms  we first compute 
using (\ref{FF1}) and (\ref{Fgamma})
\beq
 F^I \left(\hK_I {\tilde Y}_{ijk} -3 \Gamma^l_{Ii} {\tilde Y}_{ljk} \right)= 3 \mu_3 \ell^2
 e ^{\hat{K}/2} \left(\G -  k_{\hat a}\,
  G^{\hat{\bar b} \hat a}  {\bar{\CHI}}_{\hat{\bar b}}  \right) \Ox_{ijk} \ ,
\eeq
and
\beq
 F^I \partial_I   {\tilde Y}_{ijk} =
F^{z^{\ha}} \partial_{z^{\ha}}   {\tilde Y}_{ijk} = 3 \mu_3 \ell^2 k_{\hat a} \Ox_{ijk} \ ,
\label{Aagain}
\eeq
where we needed to included the linear $z^{\hat a}$ dependence
of the Yukawa couplings discussed earlier. 
When inserted into \eqref{softi}, we arrive at 
\beq \label{Aijkeff}
A_{ijk}^{\eff}= 3 \mu_3 \ell^2 e^{\hK}  \,\G \,\Omega_{ijk}  
= e^{\hK/2}\, \G \, Y_{ijk}\ .
\eeq
We see that the $A$-terms are proportional to the Yukawa
couplings.

This completes our computation of the soft terms from 
the supergravity formula \eqref{softi}.
We now compare these results to the 
terms computed in (\ref{eq:fullD3action}), (\ref{Lint})
and find agreement.

\subsection{Comparison with the D-brane action}

\subsubsection{ISD fluxes }\label{subsec:ISD}

Let us first concentrate on the simplest situation where 
only imaginary self dual (ISD) fluxes, which obey 
\eqref{ISD}, are turned on. These are the 
(2,1) and (0,3) fluxes and for this case
one has 
\beq\label{ISDcon}
\G=\CHI_{\hat a}= V_0=h_{ij}=h_{i\bj}=0 \ .
\eeq
For $\G$ and $\CHI_{\hat a}$ this can be seen directly
from the definition  \eqref{susyfromW}
while $V_0=0$ for ISD fluxes was shown in (\ref{Vhat}).
$h$ vanishes since for ISD fluxes eq.\ \eqref{e4A=alpha} holds, 
implying $h=0$ in \eqref{defh}.
Inspecting the D-brane action 
\eqref{eq:fullD3action}, \eqref{Lint} 
we see that for \eqref{ISDcon}
all soft terms
vanish and apart from the kinetic terms only
the quartic couplings survive which result in the Yukawa couplings
\eqref{tildeY}. 
Inserting \eqref{ISDcon} into the soft terms computed in the 
previous subsection we also find 
\beq
m^{2 \,\,\eff}_{i \bj,{\rm soft}}= 
B_{ij}=
m^{\eff}_g=
A_{ijk}^{\eff}=0\ .
\eeq
The supersymmetric $\mu$-terms, which we already compared
in section \ref{subsec:softD3}, also vanish for ISD fluxes. 
The vanishing of all soft terms  for (2,1) fluxes is a direct consequence
of the unbroken supersymmetry while for (0,3) fluxes 
we have instead a `strict' 
no-scale supersymmetry breaking \cite{NS} in that supersymmetry is broken
but it is not communicated to the observable sector.
Higher order $\alpha'$ and also loop corrections 
will induce soft terms and it would be interesting to study
their phenomenological properties.

\subsubsection{IASD fluxes }\label{subsec:IASD}

As we stated in section 2, ISD fluxes are well understood but 
not very interesting phenomenologically, 
as all masses and soft terms
vanish when we turn on these fluxes only. 
So in this subsection we discuss the situation where  
IASD fluxes are turned on and compare the soft terms
computed from the D-brane action with the 
supergravity formulas.

Let us start with the 
gaugino mass, which can be read off
from \eqref{Lint} to be 
\beq
m^{\Dth}= \frac14\,  e^{\hK/2} \G \ ,
\label{mgauginoD3}
\eeq
which agrees with \eqref{mshort} up to numerical factors. 

{} For the cubic terms in the D-brane action 
(\ref{eq:fullD3action}) we already accounted for the 
mixed $\phi^i  \phi^j \bar \phi^{\bk}$ (and their complex conjugate)
as arising from the effective superpotential $W^{\eff}$ as 
a mixed $\mu$-Yukawa term (c.f.\eqref{Vcubic}). This leaves 
the holomorphic trilinear terms which should be identified as $A$-terms.
Thus we have
\beq
A_{ijk}^{\Dth}= 3\mu_3 \ell^2 \,e^{\hK}\G\, \Ox_{ijk}\ (1+k_{\hat a} z^{\hat a}) \ ,
\eeq
which agrees with \eqref{Aijkeff} for  $z^{\hat a}=0$. The term linear 
in $z^{\hat a}$ could not be computed rigorously in \eqref{Aijkeff}
since one would need to know $\tilde Y_{ijk}$ in
\eqref{Aagain} to quadratic order.

For the mass terms we obtain from (\ref{eq:fullD3action}) 
\bea\label{mD3}
m_{ij}^{2(D3)} &=& \frac{36 \mu_3 \ell^2}{\Vw^2}\, h_{ij} \ ,\nn\\
m_{i\bj}^{2(D3)} &=& \frac{36 \mu_3 \ell^2}{\Vw^2}\, h_{i\bj}\ ,
\eea
which should obey
\bea\label{mcon}
m_{ij}^{2(D3)} &=& B_{ij}\ , \nn\\
m_{i\bj}^{2(D3)} &=& m_{i \bj,\, susy}^2 + m^2_{i \bj,\, {\rm soft}}
=Z^{l \bk} \mu_{il}^{\eff}  {\bar \mu}_{\bj \bk}^{\eff} 
+ m^2_{i \bj,\, {\rm soft}}\\  
&=& 6 \mu_3 \ell^2 \frac{e^{\hK}}{\Vw} 
\left[ 
-i\frac{\Vw}{w}\Gw^{\hat a \hat{b}} \, \Gw^{ \hat c \hat{d}}\; ({\bar \chi}_{\hat{b}})_{i \bar{k} \bar{l}}\,\, 
(\chi_{\hat c})_{\bj}\,^{\widetilde{\bar{k} \bar{l}}} 
\CHI_{\ha} 
\bar\CHI_{\hat{d}} + \left(\G\bar\G + \Gw^{\ha \hb} \CHI_{\ha} 
\bar\CHI_{\hbb} \right) {g}_{i \bj}\right]
\ ,\nn
\eea
where we used \eqref{muD3} and \eqref{msofteq}. 
It would be nice to confirm \eqref{mcon} more directly from the 
D-brane action. 
Here we  have determined them indirectly via a
supergravity analysis using the available input from
the action. The equations of motion, in particular the one  given in 
(\ref{e4A-alpha}), relate the trace of the mass matrix
to the IASD fluxes.  
For (3,0) fluxes we find agreement (up to numerical factors) while 
for (1,2) fluxes  the first term  
in \eqref{mcon} agrees, the second is zero in this case
but the third terms is missing. We suspect that
this is related to the problem
of not satisfying global tadpole cancellation conditions 
when turning on IASD fluxes as also noticed in ref. \cite{CIM}.

Let us close with a few phenomenological observations. 
We already observed that the $B$-terms are not proportional to the $\mu$-terms
as it is sometimes assumed in phenomenological models.
On the other hand the $A$-terms are proportional to the Yukawa couplings.
However, most importantly
the soft scalar masses \eqref{msofteq} are generically
universal as a consequence of \eqref{cancellationform} or in other words
as a consequence of the K\"ahler potential \eqref{eq:KPbrane}.

\section{Conclusions}
\label{conc}\setcounter{equation}{0}

In this paper we determined 
the low-energy effective action of type IIB string theory compactified 
on Calabi-Yau orientifolds with a stack of space-time filling D3-branes, 
in the presence
of background fluxes. Our analysis is quite general in the sense that we did not choose any particular Calabi-Yau manifold
but only demanded that it admits an isometric and holomorphic
involution. Furthermore, we considered all $h^{2,1}_- +h^{1,1}+1$ bulk
moduli surviving the orientifold projection.  
Reducing the appropriate bulk and  brane actions
we computed  the K\"ahler potential of the four-dimensional effective theory
containing the bulk moduli 
and the charged matter excitations of the brane.
In particular we determined the couplings of the complex structure
deformations to the matter fields on the brane and showed that
all fields mix non-trivially in the K\"ahler potential.
As a consequence it is generically not of the sequestered form.
In the limit of just 
one overall K\"ahler modulus and frozen
complex structure deformations, we recovered the K\"ahler
potential suggested in \cite{DWG,KKLMMT}. 
In this case the K\"ahler
potential  is of the sequestered form but
turning on the complex structure deformations already destroys this property.
  
The presence of the three-form fluxes generically
breaks supersymmetry spontaneously.
The couplings of the bulk
moduli to the matter fields on the brane communicate
this breaking to the observable sector.
Via a Giudice-Masiero mechanism `supersymmetric' fermionic masses
of the matter fields are induced. 
We computed them both directly from
the D-brane action but also from the appropriate
term in the K\"ahler potential using a supergravity analysis.
Similarly, soft $A$-terms arise which can be computed 
from the D-brane action but also 
via a supergravity relation from the Yukawa couplings.
The resulting agreement is a strong consistency check 
on our computation. For the 
soft scalar masses and the $B$-terms we can only use 
the supergravity analysis and compute these terms from
the couplings in the K\"ahler potential.

The phenomenological properties of the resulting soft terms
depend on which components of three-form flux is turned on.
For ISD fluxes one finds either no supersymmetry
breaking at all (for (2,1) fluxes) or a strict no-scale supersymmetry
breaking with all soft terms vanishing (for (0,3) fluxes).
For IASD fluxes on the other hand we find universal soft masses 
using the supergravity analysis.



It is important to note that all soft terms 
get contributions from the VEVs of the auxiliary fields in the dilaton and 
complex structure multiplets only.
Including all K\"ahler moduli does not affect the result compared to 
the situation where only one K\"ahler modulus, the overall volume, 
is taken into account.
This can be traced directly to  its ``no-scale''
property and can be seen from (\ref{Vhat}).

As already noted in \cite{CIM}, one nice feature of supersymmetry breaking by fluxes is that the supersymmetry breaking scale can be 
much smaller than the string scale. Since 3-form fluxes are quantized in units of $\ell=2 \pi \alpha'$, 
the supersymmetry breaking scale is of order $M_{susy}= \alpha'/R^3$, where $R$ is an average radius of compactification. 
In the large volume limit ($R \gg \sqrt{\alpha'}$), this is much lower than the string scale, $M_{string}= 1/\sqrt{\alpha'}$. 
Furthermore,  in the 
large radius limit, the supersymmetry breaking scale is also much lower than the masses of the Kaluza-Klein tower of states, 
$M_{KK}= 1/R$. Finally, in compactifications with a volume of order $R^6$, the 4-dimensional 
Planck mass is $M_{P}=R^3/\alpha'^2$, so it is indeed much larger than the rest of the scales in the large volume limit. 
In summary, spontaneous supersymmetry breaking by 
three-form flux in large volume compactifications leads to $M_{susy}\ll M_{KK}
\ll M_{string} \ll M_P$.     

\vskip 2 cm


\bigskip

\noindent
{\Large {\bf Appendix}}
\appendix
\renewcommand{\theequation}{\Alph{section}.\arabic{equation}}
\setcounter{equation}{0}\setcounter{section}{0}

\bigskip

\section{Kaluza-Klein reduction of Calabi-Yau orientifolds}
\label{KKreduction}

In this appendix we briefly summarize 
the calculation of the $\mathcal{N}=1$ effective action 
obtained by compactifying type IIB supergravity on Calabi-Yau 
orientifolds allowing O3/O7 planes \cite{GL}.\footnote{This computation
is performed for large Calabi-Yau manifolds, i.e. neglecting the warp
factor.} 
The spectrum of the low-energy theory 
which is invariant under the orientifold projection (\ref{o-projection})
is discussed in section \ref{spectrum}.
Inserting the expansions \eqref{transJ}--\eqref{exp1} into the
ten-dimensional type II B action one obtains (after Weyl rescaling)
\bea\label{Bulk1}
S^\text{E}_{\text{Bulk}}&=&\int_{\mathbb{M}_{3,1}} -\frac{1}{2}R*\mathbf{1}-
  \mathcal{G}_{\hat a \hat b} \; dz^{\hat a} \wedge *d\bar z^{\hat b}
  -G_{\alpha \beta} \; dv^\alpha \wedge *dv^\beta 
  - \frac{1}{4}d\, \text{ln} \cK \wedge * d\, \text{ln} \cK \nonumber \\
  &&-\frac{1}{4} d \phi \wedge * d \phi
    -\frac{\cK^2}{72} G_{\alpha \beta}\; dD_{(2)}^{\alpha}  \wedge * d D_{(2)}^{\beta} 
    -\frac{1}{8} \cK_{\alpha a b}\; dD_{(2)}^{\alpha} \wedge (c^a db^b-b^a dc^b) \nn \\
  &&  -\frac{1}{4}e^{2 \phi} dl \wedge * dl 
  -e^{- \phi } G_{ab}\; db^a \wedge * db^b
  - e^{\phi}
  G_{ab}\left(dc^a-l db^a \right) \wedge *\left(dc^b-l db^b \right)\nonumber \\
  &&-\frac{9}{8\cK^2} G^{\alpha \beta} \left(d\rho_\alpha-
  \frac{1}{2}\cK_{\alpha a b}(c^a db^b-b^a dc^b ) \right)
  \wedge
  *\left(d\rho_\beta-\frac{1}{2}\cK_{\beta cd}(c^c db^d-b^c dc^d ) \right) \nonumber \\
  && + \frac{1}{8} (\text{Im}\; \cM)^{-1\ {\hat \alpha}{\hat \beta}}
  (dU_{\hat \alpha} - (\cM dV)_{\hat \alpha})\wedge *
  (dU_{\hat \alpha} - (\bar \cM dV)_{\hat \alpha}) \ .
\eea
%
In \eqref{Bulk1} three metrics appear: $G_{\alpha \beta}(v)$ is the 
metric on $H^{(1,1)}_+$ and thus the metric for the K\"ahler deformations
$v^\alpha$.
$G_{a b}(v)$ also depends on the K\"ahler deformations but 
is the metric on $H^{(1,1)}_-$, i.e.\ the $-1$-eigenspace of $\sigma^*$.
$\mathcal{G}_{\hat a \hat b}(z)$ is the metric on the space of complex
structure deformations, i.e. a metric 
 on $H^{(2,1)}_-$. These three metrics are defined  as
\bea\label{metrics}
  G_{\alpha \beta} &\equiv& \frac{3}{2\cK}
  \int \omega_\alpha \wedge *\omega_\beta = 
  -\frac{3}{2}\left( \frac{\cK_{\alpha \beta}}{\cK}-
  \frac{3}{2}\frac{\cK_\alpha \cK_\beta}{\cK^2} \right)\ ,\nn\\
G_{a b} &\equiv& \frac{3}{2\cK}
  \int \omega_a \wedge *\omega_b = 
  -\frac{3}{2} \frac{\cK_{a b}}{\cK}\ ,\\
\mathcal{G}_{\hat a \hat b} &\equiv& 
  -\ \frac{\int \chi_{\hat a} \wedge \bar \chi_{\hat b}}
{\int\Omega\wedge\bar\Omega}\ ,\nn
\eea
where we abbreviated
\beq
\cK_{\alpha \beta} \equiv \cK_{\alpha \beta \gamma} v^\gamma\ ,
\qquad
\cK_\alpha \equiv \cK_{\alpha \beta \gamma} v^\beta v^\gamma\ ,
\qquad
\cK_{a b} \equiv \cK_{a b \alpha} v^\alpha\ ,
\eeq
and $\cK_{\alpha \beta \gamma},\cK_{a b \alpha} $ are the intersection
numbers defined in (\ref{int-num}).
The complex matrix $\cM$ is defined by using the symplectic basis
$\alpha_{\hat \alpha},\beta^{\hat \beta}$ of $H^3_+$ to be
\bea 
  \int \alpha_{\hat \alpha} \wedge * \alpha_{\hat \beta}&=&
  -(\text{Im}\; \cM +(\text{Re}\; \cM)
  (\text{Im}\; \cM)^{-1}(\text{Re}\; \cM))_{{\hat \alpha}{\hat \beta}}\ ,\nonumber \\
   \int \beta^{\hat \alpha} \wedge * \beta^{\hat \beta}&=&
   -(\text{Im}\; \cM)^{-1\ {\hat \alpha}{\hat \beta}}\ ,
  \nonumber \\
  \int \alpha_{\hat \alpha} \wedge * \beta^{\hat \beta} &=&
   -((\text{Re}\; \cM)(\text{Im}\; \cM)^{-1})_{\hat \alpha}^{\hat \beta}\ .
\eea
The self-duality $*\tilde F^{(5)} = \tilde F^{(5)}$ is ensured by
adding a term 
$\delta S^\text{E}= \frac{1}{4} dV^{\hat \alpha} \wedge dU_{\hat \alpha}
+ \frac{1}{4} dD^\alpha_{(2)} \wedge d\rho^\alpha$
to this action. The equation of motion 
for $dU_{\hat \alpha}$ and $dD^\alpha_{(2)}$ coincides with the constraints
obtained from the self-duality condition for $F^{(5)}$. 
Eliminating $dU_{\hat \alpha}$
and $dD^\alpha_{(2)}$ from the action and one obtains 
the action \eqref{N=1action_WB} for the coordinates
\eqref{tau} and the K\"ahler potential \eqref{kaehlerpot-O7-1}.
 
The same procedure can be repeated when also D-brane 
action (\ref{eq:fullD3action})
is included. In this case one finds after imposing 
the self-duality of the modified five-form field strength
for $S^E=S^\text{E}_{\text{Bulk}}+ S^\text{E}_{D3}$ 
\bea
S^\text{E} &=& \int
-\frac{1}{2}R*\mathbf{1}-
  \mathcal{G}_{\hat a \hat b} \; dz^{\hat a} \wedge *d\bar z^{\hat b}
  -G_{\alpha \beta} \; dv^\alpha \wedge *dv^\beta 
  - \frac{1}{4}d\, \text{ln} \cK \wedge * d\, \text{ln} \cK 
  -\frac{1}{4} d \phi \wedge * d \phi \nonumber \\
  &&  -\frac{1}{4}e^{2 \phi} dl \wedge * dl 
  -e^{- \phi } G_{ab}\; db^a \wedge * db^b
  - e^{\phi}
  G_{ab}\left(dc^a-l db^a \right) \wedge *\left(dc^b-l db^b \right)\nonumber \\
  &&-\frac{9}{4\cK^2} G^{\alpha \beta} \left(d\rho_\alpha-
  \frac{1}{2}\cK_{\alpha a b}(c^a db^b-b^a dc^b) -
  \mu_3\ell^2 \tr\!\!\left(\phi^i\cD_\mu\bar\phi^{\bj}
            -\bar\phi^{\bj}\cD_\mu\phi^i\right)(\omega_\alpha)_{i\bj}
            dx^\mu \right) \nn\\
  &&
  \quad \wedge
  *\left(d\rho_\beta-\frac{1}{2}\cK_{\beta cd}(c^c db^d-b^c dc^d ) -
   \mu_3\ell^2 \tr\left(\phi^i\cD_\mu\bar\phi^{\bj}
            -\bar\phi^{\bj}\cD_\mu\phi^i\right)(\omega_\alpha)_{i\bj}
            dx^\mu \right) \nn\\
  && + \frac{1}{4}\text{Im}\; \cM_{\hat \alpha \hat \beta}\; 
  dV^{\hat \alpha}\wedge *dV^{\hat \beta}
  +\frac{1}{4}\text{Re}\; \cM_{\hat \alpha \hat \beta}\;
  dV^{\hat \alpha}\wedge dV^{\hat \beta}  \nn \\
  && + \frac{\mu_3\ell^2}{2} \ 
     \big(-e^{-\phi}\tr F \wedge * F + l\,\tr\left(F\wedge F\right)\big) \nn \\
  &&+\mu_3 \ 
     \tr\left(\frac{6i\ell^2 }{\cK}\, 
            v^\alpha(\omega_\alpha)_{i\bj}
            \cD_\mu\phi^i \cD^\mu\bar\phi^{\bj} 
  +\frac{18\ell^2}{\cK^2}\,  \big(h_{i\bj}\phi^i\bar\phi^{\bj} +
  h_{ij}\phi^i\phi^j +{\rm h.c.}\big) 
  \right)*\mathbf{1} \nn \\
  &&-\mu_3 \, 
    \tr\left[ \frac{18\:\ell^2}{\cK^2}\, e^{\phi}  
        \left(g_{i\bj}g_{k\bar l} 
             \com{\phi^i}{\phi^k}\com{\bar\phi^{\bar l}}{\bar\phi^{\bj}}
         +g_{i\bj}g_{k\bar l} 
             \com{\phi^k}{\bar\phi^{\bj}}\com{\phi^i}{\bar\phi^{\bar l}} \right)\right. \nn  \\
    &&\qquad -\frac{18\ell^2e^{\phi}}{w\cK^2} \left(
     \G (1+k_{\hat a}z^{\hat a}) \Omega_{ijk}\phi^i\phi^j\phi^k 
          +3\mathcal{G}^{\hat a\hat b} 
     \CHI_{\hat a}
           \left( (1+\bar k_{\hat c}\bar z^{\hat c})
           (\bar\chi_{\hat b})_{\bar\imath\bj k}
           \bar\phi^{\bar\imath}\bar\phi^{\bj}\phi^k
          \right.\right.\nn \\
    &&\qquad\qquad\left.\left.\left.+3i
           \bar{\eta}^{\hat c}_{\hat a\hat b}\bar z^{\hat b}
           (\chi_{\hat c})_{ij\bar k}
           \phi^i\phi^j\bar\phi^k  
          + \bar \zeta_{\hat b\hat c}z^{\hat c}
           \bar\Omega_{\bar\imath\bj\bar k}
           \bar\phi^{\bar\imath}\bar\phi^{\bj}\bar\phi^{\bar k}\right)+\text{h.c.}\right)\right] *\mathbf{1}\ .  \label{BulkBrane}   
\eea
Using the coordinates given in \eqref{eq:KKc} one 
shows that the kinetic terms of \eqref{BulkBrane}  combine
into a K\"ahler metric with the
K\"ahler potential \eqref{eq:KPbrane}.

\section{Computing $F$-terms}\label{ap:F-termformulae}
In section \ref{sec:ssb} we need the $F$-terms derived from the 
K\"ahler potential $\hat K$ given in 
\eqref{kaehlerpot-O7-1} and the superpotential $\hat W$ given in \eqref{hatW}.
For convenience let us repeat these formuli here
\begin{eqnarray}\label{WK}
\hat{W}&=& \int \Omega \wedge G^{(3)}\\
 \hK(z,T,G,\tau)&=&-\text{ln}\Big[-i\int\Omega(z) \wedge \bar \Omega(\bar z) \Big] 
  -\text{ln}\big[-i(\tau - \bar \tau)\big]
  - 2 \text{ln}\Big[\frac16\cK(T,G,\tau)\Big]\ ,\nn
  \label{kaehlerpot}
\end{eqnarray}
where 
$\cK \equiv \cK_{\alpha \beta \gamma} v^\alpha v^\beta v^\gamma$.
The $v^\alpha$ depend implicitly via the equation 
\beq
  T_\alpha = \frac{3i}{2} \rho_\alpha 
  + \frac{3}{4}\cK_{\alpha} 
- \frac{3i}{4(\tau-\bar \tau)}\, \cK_{\alpha b c}G^b (G- \bar G)^c \ , 
\eeq
on the K\"ahler coordinates $T,G$ and $\tau$ defined as
\beq
 \tau =l+i e^{- \phi}\ , \qquad
  G^a = c^a -\tau b^a\ ,
\eeq
and we abbreviated 
$\cK_{\alpha} \equiv \cK_{\alpha \beta \gamma} v^\beta v^\gamma$.
The derivatives of the K{\"a}hler potential $K_I \equiv \partial_I K$
are 
\beq\label{Kfirst}
  {\hat K}_{\tau}       = \frac{i}{2}e^{\phi} +i G_{ab}b^a b^b \ ,\qquad
  {\hat K}_{T_{\alpha}} = -2\frac{v^\alpha}{\cK}  \ ,\qquad
  {\hat K}_{G^a}=2i G_{ab}b^b   \ ,\qquad
\eeq
where $G_{ab}$ is defined in \eqref{metrics}.
For the complex structure one has (c.f.\ eq.\ \eqref{derivomega})
\cite{Candelas:1990pi}
\beq
{\hat K}_{z^{\hat a}} = -k_{\hat a}\ , \qquad 
\partial_{\hat a} \Omega = k_{\hat a} \Omega + i \chi_{\hat a}\ .
\eeq

The K\"ahler metric is derived by taking one more (antiholomorphic)
derivative leading to
\bea\label{Kmetric}
 \hat K_{T_\alpha \bar T_\beta} &=& \frac{G^{\alpha \beta}}{\cK^2}\ ,\nn
\\
  \hat K_{T_\alpha \bar G^a} &=& -\frac{3i}{4} \frac{G^{\alpha
\beta}}{\cK^2} \cK_{\beta a b} b^b\ , \nn\\
  \hat K_{T_\alpha \bar \tau} &=& -\frac{3i}{4} \frac{G^{\alpha
\beta}}{\cK^2} \cK_{\beta a b} b^a b^b\ , \nn\\
  \hat K_{G^a \bar G^b} &=& e^\phi G_{ab}+\frac{9}{4} \frac{G^{\alpha
\beta}}{\cK^2} \cK_{\beta a c} b^c
   \cK_{\beta b d} b^d\ , \\
  \hat K_{G^a \bar \tau} &=& e^\phi G_{ab}b^b+\frac{9}{8} \frac{G^{\alpha
\beta}}{\cK^2} \cK_{\beta a c} b^c
   \cK_{\beta b d} b^b b^d\ , \nn\\
   \hat K_{\tau \bar \tau} &=& \frac{1}{4} e^{2\phi} +
   e^\phi G_{ab}b^a b^b +\frac{9}{16} \frac{G^{\alpha \beta}}{\cK^2}
\cK_{\beta a c} b^a b^c
   \cK_{\beta b d} b^b b^d\ , \nn\\
   \hat K_{z^{\hat a} \bar z^{\hat b}} &=& \mathcal{G}_{\hat a \hat b}\ .\nn
\eea
Its inverse is found to be
\bea\label{Kinvers}
  \hat K^{T_\alpha \bar T_\beta} &=& \cK^2 G_{\alpha \beta} - \frac{9}{4}
e^{-\phi} G^{ab} \cK_{\beta a c} b^c
   \cK_{\beta b d} b^d + \frac{9}{4} e^{-2\phi} \cK_{\beta a c} b^a b^c
   \cK_{\beta b d} b^b b^d\ , \nn\\
  \hat K^{T_\alpha \bar G^a}&=&-\frac{3i}{2} e^{-\phi} G^{ab} \cK_{\alpha
b c}b^c
  - 3i e^{-2\phi}\cK_{\alpha b c} b^b b^c b^a\ , \nn \\
  \hat K^{T_\alpha \bar \tau}&=&3i e^{-2\phi} \cK_{\alpha b c}b^c \ , \nn
\\
  \hat K^{G^a \bar G^b} & = &e^{-\phi} G^{ab} + 4 e^{-2 \phi} b^a b^b\ ,
\nn \\
  \hat K^{G^a \bar \tau} & = &-4 e^{-2 \phi} b^a \\
  \hat K^{\tau \bar \tau} & = & 4e^{-2\phi}\ , \nn \\
  \hat K^{z^{\hat a}\bar z^{\hat b}} &=&\mathcal{G}^{\hat a \hat b} \nn\ .
\eea

Using \eqref{Kfirst} and  \eqref{WK} we can determine  the 
covariant derivatives $D_I W \equiv (\partial_I+ K_I )W$ to be
\bea
  D_{\tau}{\hat W}&=& \frac{i}{2}e^{\phi} \bar\G
             +i G_{ab}b^a b^b\ {\hat W} \ , \qquad 
  D_{T_{\alpha}}{\hat W}\ =-2\ \frac{v^\alpha}{\cK}\ {\hat W} \ , \nn\\
  D_{G^a}{\hat W}&=&2i G_{ab}b^b\ {\hat W}\ , \qquad 
  D_{z^{\hat a}}{\hat W}\ =\ \CHI_{\hat a} \ ,
  \label{kcov1}
\eea
where 
\beq
\G \equiv \int {\bar \Omega} \wedge G^{(3)}\ , 
\qquad \CHI_{\ha}\equiv \int \chi_{\ha} \wedge G^{(3)}\ .
\label{Idef}
\eeq
Using \eqref{Kmetric} and \eqref{kcov1} one computes the 
F-terms needed in the main text.
They are defined as  $F^I = e^{{\hat K}/2} K^{I \bar J} D_{\bar J}{\hat W}$ 
and read
\begin{eqnarray}\label{Fterms}
 F^{\tau} &=& - 2i e^{{\hat K}/2}  e^{-  \phi} \G\ , \nonumber \\
 F^{T_\alpha} &=& e^{{\hat K}/2}\left(-\frac{3}{2} \cK_\alpha \hat{\bar  W} + 
                  \frac{3}{2}e^{-\hat \phi} 
 \cK_{\alpha a b} b^a b^b \G \right), \nonumber \\
 F^{G^a} &=& e^{K/2} 2i e^{-  \phi} b^a \G \ ,\\
F^{z^{\hat a}}&=& e^{{\hat K}/2}
  {\cal G}^{\hat a \hat{b}} \bar{\CHI}_{\hat{b}} \  .\nn
 \label{F^A}
\end{eqnarray}
Using \eqref{Fterms} and \eqref{Kfirst} one obtains the 
 following useful identities
\begin{eqnarray}\label{usefull}
F^I\hat{K}_I&=& e ^{\hat{K}/2} \left(\G -  k_{\hat a}\,
  {\cal G}^{\hat{b} \hat a}  {\bar{\CHI}}_{\hat {b}}   + 3\,
    \hat{\bar W} \right) \nn \\
  F^I \partial_I v^\alpha &=& - \frac{1}{2} e^{{\hat K}/2} v^\alpha \hat{\bar W},\ \ \ \ \ \
  F^I \partial_I \cK = -\frac{3}{2} e^{{\hat K}/2} \cK \hat{\bar W},
\\
 F^I F^{\bar J} \partial_I \partial_{\bar J} v^\alpha &=& 
  -\frac{1}{4} e^{\hK} |{\hW}|^2 v^\alpha, \ \ \ \ \ \
  F^I F^{\bar J} \partial_I \partial_{\bar J} \cK = -\frac{3}{4} e^{\hK} |\hW|^2 \cK. \nn
  \label{FF1}
\end{eqnarray}

The other quantities we need in the main text are the 
the Christoffel following symbols defined as
$\Gamma^k_{i I} = Z^{k \bar j} \partial_I Z_{\bar j i}$ for the 
matter metric 
\beq
Z_{i \bar j} =  -i\frac{6\mu_3\ell^2}{\cK}\, v^\alpha (\omega_\alpha)_{i\bar j}\ .  
\eeq
One finds
\bea
  \Gamma^k_{i\tau} &=& \frac{3\mu\ell^2}{2 \cK} 
\kappa^{\alpha \beta} 
\cK_{\beta a b} b^a b^b 
                     (\omega_\alpha)_{\bar j i} Z^{k \bar j} 
                 -\frac{i}{2}  G_{ a b} b^a b^b \delta_{k}^{i}\ ,\nn\\
  \Gamma^k_{i G^a} &=& \frac{3 \mu_3\ell^2}{\cK} \kappa^{\alpha \beta} \kappa_{\beta a b} b^b 
                     (\omega_\alpha)_{\bar j i} Z^{k \bar j} 
                    - i  G_{a b} b^b \delta_{k}^{i}\ ,\\
  \Gamma^k_{i T^\alpha} &=& 
      -    \frac{2i\mu_3\ell^2}{\cK} \kappa^{\alpha \beta} (\omega_\beta)_{\bar j i} Z^{k \bar j} 
                    + \frac{1}{\cK} v^\alpha \delta_{k}^{i} \ ,\nn
\label{Christoffel}
\eea
obeying the identity
\beq \label{Fgamma}
F^I\Gamma^l_{Ii}= e ^{\hat
  K/2} \hat{\bar W} \delta ^l_i . 
\eeq

\section{Normal coordinates expansion}
\label{pullback}
In the Dirac-Born-Infeld action \eqref{eq:DBIaction} and the Chern-Simons action \eqref{eq:ChernSimon} of the D$p$-brane, there appear various contributions which have to be pulled back from the space-time manifold $M$ to the world-volume $\WV$ of the brane via $\varphi:\WV\hookrightarrow M$. 

The pull-back of the D-brane action contains the whole dynamics of the brane in the following way: The pull-back is not just performed with a rigid map $\varphi$, but we allow for small fluctuations normal to the embedded world-volume $\WV\subset M$. Hence we must describe these fluctuations in an appropriate way. 

The fluctuations of the embedded world-volume are described by considering displacements of the embedding in the normal direction of the world-volume as in \cite{Friedan:1980jm,Alvarez-Gaume:hn}. These displacements are encoded in sections of the normal bundle $N\WV$ of a fixed world-volume. Let us take the section $\xi$ of the normal bundle to represent a fluctuation. The section $\xi$ gives rise to a map $\hat\varphi:\WV\times I \hookrightarrow M,(y,t)\mapsto \hat\varphi(y,t)$ with the following properties:
\begin{equation}
   \hat\varphi(y,0)=\varphi(y)  \  ,
\end{equation}
and for fixed $y$ the function $\hat\varphi(y,t)$ parameterizes a geodesic in the direction $\xi|_y$ in such a way that the geodesic from $t=0$ to $t=1$ has arc-length $\norm{\xi}$. In mathematical terms the function $\hat\varphi$ is given by the exponential map of $\xi$. Thus
\begin{equation}
   \frac{d}{dt}\hat\varphi(y,0)=\xi|_y \  ,
\end{equation}
and we extend $\xi$ to $\WV\times I$ by parallel transport along t. As there exists a tubular neighborhood of $\WV$, we always have this construction for sufficiently small fluctuations $\xi$.

Having expressed the fluctuations in terms of the map $\hat\varphi$, we are now able to pull-back a tensor $T$ of $M$ and expand using
\begin{align}
   \left.\left(\hat\varphi^*T\right)\right|_{\hat\varphi(y,t)} &= 
     \hat\varphi^*\left.\left(e^{\nabla_{t\xi}}T\right)
                  \right|_{\hat\varphi(y,0)} \nonumber \\
     &=\hat\varphi^*\left.\left(T\right)\right|_{\hat\varphi(y,0)}+
       t\hat\varphi^*\left.\left(\nabla_\xi T\right)\right|_{\hat\varphi(y,0)}+
       \frac{1}{2}t^2\hat\varphi^*\left.\left(\nabla_\xi\nabla_\xi T
                                  \right)\right|_{\hat\varphi(y,0)}+\ldots  \  .
\end{align}

Recall that the Riemann tensor and the torsion of a (Pseudo--) Riemannian manifold is given by
\begin{align}
   R(X,Y)Z&=\nabla_X\nabla_YZ-\nabla_Y\nabla_XZ-\nabla_{\lie{X}{Y}}Z 
      \label{eq:riemann} \\
   T(X,Y)&=\nabla_XY-\nabla_YX-\lie{X}{Y}=0 \label{eq:torsion} \ ,
\end{align}
where the Levi-Cevita connection $\nabla$ is metric and torsion-free. 

In local coordinates $(x^\mu,t)$ of $U\times I\subset \WV\times I$ where $\mu=1,\ldots,\dim\WV$, we have the vector fields $(\partial_\mu,\partial_t)$. Since the Lie-bracket $\lie{\partial_t}{\partial_\mu}$ vanishes and we get \eqref{eq:torsion}
\begin{equation}
   \nabla_\xi \partial_\mu = \nabla_{\partial_\mu} \xi  \ ,
\end{equation}
and hence with \eqref{eq:riemann} and $\nabla_\xi \xi=0$
\begin{equation}
   R(\xi,\partial_\mu)\xi=\nabla_\xi\nabla_{\partial_\mu}\xi
      =\nabla_\xi\nabla_\xi \partial_\mu \ .
\end{equation}
The developed tools we readily apply to the metric and obtain up to second order in the parameter $t$
\begin{equation}
\begin{split}
   \left.\left(\hat\varphi^*g(\partial_\mu,\partial_\nu)\right)
   \right|_{\hat\varphi(y,t)}  
     =&\hat\varphi^*\left.\left(g(\partial_\mu,\partial_\nu)\right)
                    \right|_{\hat\varphi(y,0)}+\\
      &+t\hat\varphi^*\left.\left(g(\nabla_{\partial_\mu }\xi,\partial_\nu )\right)
                     \right|_{\hat\varphi(y,0)}
      +t\hat\varphi^*\left.\left(g(\partial_\mu ,\nabla_{\partial_\nu }\xi)\right)
                     \right|_{\hat\varphi(y,0)} \\
      &+t^2\hat\varphi^*\left.\left(g(\nabla_{\partial_\mu}\xi,\nabla_{\partial_\nu}\xi)
                              \right)\right|_{\hat\varphi(y,0)} 
       +t^2\hat\varphi^*\left.\left(g(R(\xi,\partial_\mu )\xi,\partial_\nu )\right)
                              \right|_{\hat\varphi(y,0)}+\ldots \ .
\end{split}
\end{equation}

This index free notation translates in a slightly abusive way into the component expression (for $t=1$)
\begin{equation} \label{eq:gpbcomp}
\begin{split}
   \varphi^*(g)_{\mu\nu}=&g_{\mu\nu}+\ell g_{\mu n}\D_\nu\phi^n+
         \ell g_{\nu n}\D_\mu\phi^n \\
     &+\ell^2 g_{nm}\D_\mu\phi^n\D_\nu\phi^m+
         \ell^2 g_{\mu\tau}\Riem{n}{\tau}{\nu}{m}\phi^n\phi^m +\ldots \ ,
\end{split}
\end{equation}
where the $\ell\phi$s denote the fluctuations of the brane in the normal direction and $\D$ is a covariant derivative of the normal bundle. Greek indices are used for the world-volume coordinates of the brane and Latin indices for the normal directions of the world-volume. As we consider a space-time filling D3-brane, the Greek indices correspond also to the non-compact space-time manifold and the Latin indices to the internal (Calabi-Yau) manifold. In the same way one can derive the pull-back of various forms appearing in \eqref{eq:DBIaction} and \eqref{eq:ChernSimon}. For a $p$-form we obtain in local coordinates up to second order (for $t=1$)
\begin{equation} \label{eq:ppbcomp}
\begin{split}
   \left.\left(\hat\varphi^*C^{(p)}\right) \right|_{\hat\varphi(y,t)} 
   =&\left(\frac{1}{p!}C^{(p)}_{\nu_1\ldots\nu_p} 
    +\frac{\ell}{p!}\phi^n\partial_n(C^{(p)}_{\nu_1\ldots\nu_p})
    -\frac{\ell}{(p-1)!}\D_{\nu_1}\phi^n C^{(p)}_{n\nu_2\ldots\nu_p} \right. \\
   &+\frac{\ell^2}{2p!}\phi^n\partial_n(\phi^m\partial_m(C^{(p)}_{\nu_1\ldots\nu_p}))
    -\frac{\ell^2}{(p-1)!}\D_{\nu_1}\phi^n
    \cdot\phi^m\partial_m(C^{(p)}_{n\nu_2\ldots\nu_p}) \\
   &+\frac{\ell^2}{2(p-2)!}\D_{\nu_1}\phi^n\D_{\nu_2}\phi^m C^{(p)}_{nm\nu_3\ldots\nu_p}\\
   &\left.+\frac{p-2}{2p!}\ell^2\Riem{n}{\tau}{\nu_1}{m}\phi^n\phi^m
    C^{(p)}_{\tau\nu_2\ldots\nu_p}\right)
    dx^{\nu_1}\wedge\ldots\wedge dx^{\nu_p}  \ .
\end{split}
\end{equation}

\section{Decomposition of the 10D spinor.}\label{ap:spinor}

The spinor $\theta$ is a 10d Majorana-Weyl fermion of negative chirality, i.e. in the ${\bf 16}$ of
$SO(9,1)$. Under $SO(9,1)\rightarrow SO(3,1) \times SO(6)$, it decomposes into $({\bar {\bf 2}}, {\bf 4}) 
\oplus ( {\bf 2}, {\bar {\bf 4}})$. From the 4-dimensional point of view there are
four fermions, the gaugino $\lambda$, superpartner of the gauge field $A_{\mu}$, and 3 fermions $\psi^i$,superpartners 
of the scalars $\phi^i$.
The fermions $\psi^i$ and $\lambda$ should then be contained in $\theta$, in the following way
\beq
\theta= a \,\psi^i \otimes \chi_i + b\, \lambda \otimes \chi_{4} 
+\, {\rm h.c.} \ ,
\label{decomp}
\eeq
where $i=1,2,3$ and $a$ and $b$ could be functions of internal space.
In order to find these functions, and the precise form of $\chi_i$ and $\chi_{4}$, let us perform a world-volume supersymmetry 
transformation of the world-volume fields $\phi^i, A_{\mu}$, and compare it to the standard ${\mathcal N}=1$ supersymmetry 
transformation of the scalar in a chiral and the vector in a vector multiplet. World-volume fields transform according to
\bea
\delta_{\epsilon} \phi^i&=&{\bar \epsilon}\, {\tilde \Gamma}^i \theta  \label{wvsusyscalar}\ ,\\
\delta_{\epsilon} A_{\mu}&=&{\bar \epsilon}\, \Gamma_{\mu} \theta\ ,
\label{wvsusyA}  
\eea
where ${\tilde \Gamma}^i$ is a gamma-matrix for the warped metric.
On the other hand, the usual ${\mathcal N}=1$ supersymmetry transformations of the scalar in 
a chiral multiplet and the vector in 
the vector multiplet are
\bea
\delta_{\epsilon} \phi^i&=&\epsilon \psi^i   \label{stsusyscalar}\ ,\\
\delta_{\epsilon} A_{\mu}&=&{\bar \epsilon}\, \gamma_{\mu} \lambda \ .
\label{stsusyA}  
\eea
We want to perform a supersymmetry transformation that leaves the supergravity background invariant. With a background metric
of the warped form, eq.(\ref{metric}),  the invariant spinor is \cite{GP}
\beq
\epsilon=e^{A/2} \left(\epsilon_4 \otimes \chi + \epsilon_4^* \otimes \chi^*\right)\ ,
\label{susy}
\eeq
where $\epsilon_4$ is a constant spinor in 4 dimensions with positive chirality, $\gamma_{5}\epsilon_4=\epsilon_4$,
and $\chi$ is a nowhere vanishing covariantly constant with respect to the unwarped metric $g_{mn}$, i.e., $\D_m \chi =0$, and has negative 6d chirality.
We can take $\chi$ to be normalized to one, i.e. $\chi^{\dagger}\chi=1$.
By doing an $SO(6)$ rotation, we can also choose $\chi$ to be annihilated by all lowering operators 
\beq
\gamma^{\ib} \chi =0.
\eeq
Using the formula for the world-volume 
supersymmetry transformations (\ref{wvsusyA}), 
for a supersymmetry parameter given by (\ref{susy}),
 and inserting 
the decomposition (\ref{decomp}) we get
\beq
\delta_{\epsilon} A_{\mu}= b\,e^{A/2}\,{\bar \epsilon_4} \gamma_{\mu} \lambda
\, \, \chi^{\dagger} \chi_{4}
= b\,e^{3A/2} \,{\bar \epsilon_4} {\tilde \gamma}_{\mu} \lambda
\,\,  \chi^{\dagger} \chi_{4}.
\eeq
Then, taking $b=e^{-3A/2}$; $\chi_{4}=\chi$, we recover the
global ${\mathcal N}=1$ supersymmetry transformation (\ref{stsusyA}).
Repeating the process for $\phi^i$, we get
\beq
\delta_{\epsilon} \phi^i= a\,e^{3A/2} \,\epsilon_4 \psi^j \,  \chi^{T} {\gamma}^i \chi_j\ .
\eeq
We recover the usual ${\mathcal N}=1$ supersymmetry transformation for the scalar in the chiral multiplet if we set
\beq
a \,\chi_i= \frac{1}{3! \|\Omega\|}\,e^{-3A/2} \Omega_{ijk} \gamma^{jk} \chi
\eeq
where the $\gamma$s are taken with respect to the unwarped metric
$g_{mn}(y)$, 
$\Omega$ is the holomorphic (3,0)-form of the Calabi-Yau,  
$\|\Omega\|^2=\frac{1}{3!}\Omega_{ijk} {\bar \Omega}^{ijk}$ 
and we have used
\beq
\chi^T \gamma^{ijk} \chi= \frac{1}{\|\Omega\|}{\bar \Omega}^{ijk}. 
\label{omega}
\eeq

Combining both results, the spinor $\theta$ is
\beq
\theta= e^{-3A/2}\,\left(\frac{1}{6 \|\Omega\|}\,\Omega_{ijk}\, \psi^i \otimes
{\gamma}^{jk}\, \chi 
+  \lambda \otimes \chi \right) + h.c.
\eeq

When inserting this spinor in the fermionic Lagrangian (\ref{fermlagrangian}), we get a kinetic term of the form 
\beq
{\cal L}_{kin}= -i \mu_3 \frac{\sqrt 6}{\Vw^{3/2}} Tr \left(\frac{2}{3} \overline{\psi}^{\ib}{ \gamma}^{\mu} 
D_{\mu} \psi^l {\tilde g}_{\ib l} +6 \overline{\lambda}
{\gamma}^{{\mu}} 
D_{\mu} \lambda \right) .\label{Lkin1}
\eeq

From supersymmetry, the fermionic and bosonic kinetic terms should have a common K{\"a}ler metric, i.e.
\bea
{\cal L}_{kin,susy}&=&  -K_{\ib l}
  D_{\mu} \overline{\phi}^{\ib} 
D^{\mu} \phi^l-i K_{\ib l}  \overline{\psi'}^{\ib}{\gamma}^{\mu} 
D_{\mu} \psi'^l \nn \\
&&-Re \left[\frac{1}{4}f_{ab} (F^a + i*_4 F^a)_{\mu \nu} (F^b+i*_4F^b) ^{\mu \nu} + i f_{ab} {\bar \lambda} ^a {\gamma}^{\hat{\mu}} 
D_{\mu} \lambda ^b  \right]\, ,
\label{Lsusy}
\eea
where $K_{\ib l}$ is the derivative of the K{\"a}hler potential. 

From the bosonic DBI action (\ref{eq:DBIlow}), we see that 
$K_{\ib l}=\frac{6\ell^2}{\Vw} g_{\ib
l}$. Then, from susy, we should have the same metric in the
kinetic term for the fermions in the chiral multiplet, which is
different from the $\Vw^{-3/2} g_{\ib
l}$ that we have in (\ref{Lkin1}) \footnote{If we had found the complete action (bosonic and fermionic) from the
supersymmetric DBI-CS, supersymmetry would be automatic. But since we need bosonic terms that are appear only in the
non-Abelian action, and we do not have its supersymmetric version, we are forced to compute bosonic and fermionic terms separately.}  
. This means
that we need to renormalize the fermions $\psi^i$, i.e. the actual susy
partners of the bosons
$\phi ^i$ are the fermions $\psi'\vphantom{\psi'} ^i$, defined as
\beq
\psi'^i= \frac{1}{\ell}\,\left(\frac{2}{27 \,\Vw}\right)^{1/4} \psi^i.
\eeq
Besides, we see from the bosonic DBI plus CS action (\ref{eq:fullD3action}) that
\beq
f_{ab}=-i \mu_3 \ell^2 \tau \delta_{ab}
\label{fab}
\eeq
which means that we also need to renormalize the gaugino, such that its kinetic term follows the
form (\ref{Lsusy}), namely
\beq
\lambda'= \frac{1}{\ell} \,\left(\frac{6}{\Vw}\right)^{3/4} \lambda.
\eeq
Then, the decomposition of $\theta$ in terms of these spinors is
\beq
\theta= \ell \,e^{-3A/2}\,\left(
  \left(\frac{\Vw}{6}\right)^{1/4}\frac{1}{2 \|\Omega\|}
\,\Omega_{ijk}\, \psi'^i \otimes
{\gamma}^{jk}\, \chi 
+ \left(\frac{\Vw}{6}\right)^{3/4} e^{-\hphi/2}\,  \lambda' \otimes \chi \right) + {\rm h.c.}
\eeq
and kinetic term for the fermions is of the form (\ref{Lsusy}). In the main sections of the paper we have dropped the primes 
for the fermions.

\vskip 1cm

\subsection*{Acknowledgments}

This work is supported by DFG -- The German Science Foundation,
GIF -- the German--Israeli Foundation for Scientific Research,
the European RTN Program HPRN-CT-2000-00148, the
DAAD -- the German Academic Exchange Service. The work of M.G. 
is also partially supported by EEC contracts
HPRN-CT-2000-00122, HPRN-CT-2000-00131.

We have greatly benefited from conversations and correspondence with 
J. Distler,
E. Dudas, A. Frey, E.Kiritsis, A. Micu,
R. Minasian, A. Tomasiello and A. Uranga.


\providecommand{\href}[2]{#2}
\end{document}